\documentclass[runningheads]{llncs}

\usepackage{eccv}

\usepackage{eccvabbrv}
\usepackage{multirow}
\usepackage{graphicx}
\usepackage{booktabs}
\usepackage{colortbl}
\usepackage{array}
\usepackage[accsupp]{axessibility}  % Improves PDF readability for those with disabilities.

\usepackage{hyperref}

\usepackage{orcidlink}
\definecolor{cvprblue}{rgb}{0.21,0.49,0.74}
\definecolor{lightskyblue}{rgb}{0.53, 0.81, 0.98}
\definecolor{lightblue}{rgb}{0.68, 0.85, 0.9}
\definecolor{softblue}{rgb}{0.85, 0.91, 0.98}
\definecolor{mygray}{gray}{0.65}

\begin{document}

\title{\textit{DiffuX2CT}: Diffusion Learning to Reconstruct CT Images from Biplanar X-Rays} 

\titlerunning{DiffuX2CT}

\author{Xuhui Liu\inst{1} \and Zhi Qiao \inst{2} \and Runkun Liu \inst{2} \and Hong Li  \inst{1} \and Juan Zhang$^*$  \inst{1} \and \\
Xiantong Zhen$^*$ \inst{2} \and  Zhen Qian \inst{2} \and Baochang Zhang \inst{1,3} }

\authorrunning{X. Liu et al.}

\institute{Beihang University, Beijing, China \and
Central Research Institute, United Imaging Healthcare, Beijing, China\\ \and
Zhongguancun Laboratory, Beijing, China\\
\email{\{xhliu,zhang\_juan\}@buaa.edu.cn}, \email{zhenxt@gmail.com}}

\maketitle

\renewcommand{\thefootnote}{\fnsymbol{footnote}}
\footnotetext[1]{Corresponding authors.}
\begin{abstract}
Computed tomography (CT) is widely utilized in clinical settings because it delivers detailed 3D images of the human body. However, performing CT scans is not always feasible due to radiation exposure and limitations in certain surgical environments. As an alternative, reconstructing CT images from ultra-sparse X-rays offers a valuable solution and has gained significant interest in scientific research and medical applications. However, it presents great challenges as it is inherently an ill-posed problem, often compromised by artifacts resulting from overlapping structures in X-ray images. In this paper, we propose \textit{DiffuX2CT}, which models CT reconstruction from orthogonal biplanar X-rays as a conditional diffusion process. \textit{DiffuX2CT} is established with a 3D global coherence denoising model with a new, implicit conditioning mechanism. We realize the conditioning mechanism by a newly designed tri-plane decoupling generator and an implicit neural decoder. By doing so, \textit{DiffuX2CT} achieves structure-controllable reconstruction, which enables 3D structural information to be recovered from 2D X-rays, therefore producing faithful textures in CT images. As an extra contribution, we collect a real-world lumbar CT dataset, called LumbarV, as a new benchmark to verify the clinical significance and performance of CT reconstruction from X-rays. Extensive experiments on this dataset and three more publicly available datasets demonstrate the effectiveness of our proposal.
\keywords{CT Reconstruction from Biplanar X-rays \and Implicit Conditioning Mechanism \and Tri-plane Representation}
\end{abstract}

\section{Introduction}
\label{sec:intro}

\begin{figure*}[t]
\begin{center}
\includegraphics[width=0.99\textwidth]{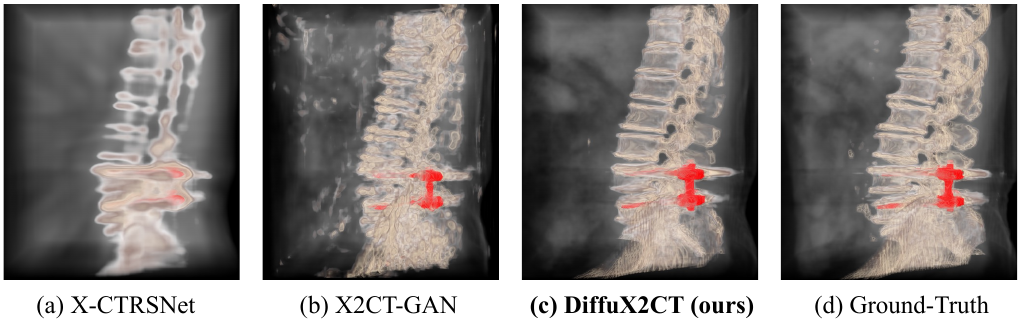}
\end{center}
   \caption{Comparison of different reconstruction methods. The reconstructed CT images are rendered for a more intuitive illustration. Our \textit{DiffuX2CT} significantly outperforms competitors including both generative and regression methods. \textit{DiffuX2CT}  generates high-quality and faithful CT images. Notably, it recovers precise bone structures and correctly positioned implants, which are close to ground truth.}
\label{fig:intro}
\end{figure*}

Computed tomography (CT) is one of the widely used imaging modalities in clinical routines. CT images provide detailed and comprehensive views of target areas of interest, thereby playing a crucial role in clinical practice for diagnosis \cite{wang2018image}. To acquire high-quality CT images, a full rotational scan with numerous X-ray projections is performed around the body. Standard reconstruction algorithms, such as filtered back projection and iterative reconstruction, facilitate the creation of these volumes \cite{herman2009fundamentals}. This process allows physicians to accurately visualize the location and shape of structures like bones and implants \cite{siasios2017percutaneous}. However, the extensive data acquisition in CT scans makes it infeasible in most surgical scenarios. In addition, it is also less preferred to perform CT scans due to the higher radiation exposure to patients compared to a single X-ray.
Fortunately, reconstruction of CT images from ultra-sparse X-rays offers an attractive alternative \cite{shen2019patient}, which only requires cost-effective X-ray machines and reduces radiation exposure to patients. 
While X-rays excel in bone contrast and enable rapid assessments \cite{khan2019comparing}, they only project objects onto a 2D plane, lacking a complete 3D view of internal structures. Specifically, the limited internal body information in X-rays leads to significant ambiguity and artifacts, as numerous possible CT images could be obtained from the same set of 2D X-ray projections. In addition, the overlapping of objects in X-rays further complicates the reconstruction of CT images. 
  
Due to the significance in both research and clinical practice, great efforts have been made in 3D CT reconstruction from X-rays \cite{shen2019patient,shiode20212d,ying2019x2ct,henzler2018single,wu2021reconstruction,kasten2020end,ge2022x,jiang2021reconstruction,zhang2023xtransct}.
% Some researchers turn to utilizing image registration as an alternative \cite{}.
Early attempts treat it as a regression task, which usually utilizes deep neural networks \cite{ronneberger2015u} to directly establish a mapping from X-rays to CT images. However, these methods tend to generate blurry CT images, failing to recover realistic body structures (Fig.~\ref{fig:intro}(a)). This is largely due to regression losses that simply calculate the averaged results of possible predictions. Deep generative models, e.g., generative adversarial networks (GANs) \cite{goodfellow2014generative}, have also been explored, which perform better in learning the distribution of body anatomy from a large training set, thereby delivering better sample quality \cite{ying2019x2ct,jiang2021reconstruction}. However, these GAN-based methods fall short of recovering sufficient 3D structure from 2D X-rays, yielding severe artifacts and blurry images (Fig.~\ref{fig:intro}(b)).
% merely employ naive operations to extract X-ray features, such as duplicating to improve data dimensions and directly adding diverse view features, 
Recently, diffusion models \cite{ho2020denoising,sohl2015deep} have achieved tremendous success in various synthesis tasks~\cite{song2020denoising,dhariwal2021diffusion,vahdat2021score,zeng2023face,rombach2022high,liu2024ladiffgan,peebles2022scalable,li2024uv}. They have also been tried on settings of sparse-view and limited-angle CT reconstruction problems~\cite{chung2023solving,lee2023improving,liu2023dolce}. 
Unlike the challenging setting addressed in our work, they still rely on CT scanners, not offering a solution for the extremely ill-posed problem of using biplanar X-rays in CT reconstruction. Moreover, they are developed under 2D network architectures, lacking the ability to establish 3D global coherence in CT images.

This paper presents a new 3D conditional diffusion model, termed \textit{DiffuX2CT}, for CT reconstruction from biplanar orthogonal X-rays. This is different from current diffusion-based CT reconstruction methods, which still rely on CT scans. At a high level, \textit{DiffuX2CT} formulates the reconstruction task as a conditional denoising diffusion process, where biplanar orthogonal X-rays are incorporated as a condition in the diffusion process. \textit{DiffuX2CT} is implemented with a 3D encoder-decoder network to model the 3D prior distribution of body anatomy, ensuring high sample quality and global coherence throughout the 3D CT image. To reduce computation costs, \textit{DiffuX2CT} introduces the shifted-window attention mechanism \cite{liu2022video} as a substitute for the self-attention \cite{vaswani2017attention,arnab2021vivit} operations. 
Moreover, we propose a new, implicit conditioning mechanism (ICM) to incorporate 3D structure information into CT image generation in the diffusion process. The ICM is realized by a newly designed tri-plane decoupling generator and an implicit neural decoder.
The tri-plane decoupling generator individually reconstructs the three 2D planes of the tri-plane representations \cite{chan2022efficient}, while preserving their spatial correlations, enabling the effective recovery of 3D structure from 2D biplanar X-rays. 
% To achieve this, it first employs an interleaved convolution to enhance the correlated features of biplanar X-rays mutually and then extracts the corresponding two feature planes. Meanwhile, it defines two view modulators to combine them adaptively and integrates a learnable embedding to establish the third feature plane. 
Subsequently, the implicit neural decoder maps the tri-plane features and the assigned coordinates to 3D implicit representations for conditioning the synthesis process. 
% }
% The implicit conditioning network can also work in a regression manner, which we call \textit{DiffuX2CT-Reg}.
% (Fig.~\ref{fig:intro}(c)).}
% The implicit conditioning network is able to recover 3D structure from 2D biplanar X-rays, which can also work with regression models (Fig.~\ref{fig:intro}(c)).
In addition, we introduce a geometry projection loss in training \textit{DiffuX2CT} to facilitate structural consistency in the 3D space. 

To verify the clinical significance of CT reconstruction using biplanar X-rays, we collect a new real-world lumbar vertebra dataset, called LumbarV, with 268 3D CT images from different patients. Each contains implants inserted in the lumbar vertebrae. The \textit{DiffuX2CT} is capable of reconstructing high-quality CT images, which could assist surgeons in locating the implants and measuring the shape of the vertebrae as shown in Fig.~\ref{fig:intro}(c).

The major contributions of our work are summarized in three folds as follows:
\begin{itemize}
\item We propose a new, 3D conditional diffusion model - \textit{DiffuX2CT}, which is the first model that leverages diffusion models with a new tri-plane conditioning mechanism to achieve faithful and high-quality CT image reconstruction from biplanar X-rays.
% This is the first work based on the diffusion model for faithful CT reconstruction from biplanar X-rays.

\item We propose the implicit conditioning mechanism (ICM), which is realized by a newly designed tri-plane decoupling generator and an implicit neural decoder. Our implicit conditioning mechanism enables 3D structure information to be recovered for reconstructing CT images.

\item We contribute a new real-world lumbar vertebra dataset,
% \zqiao{in which post-operative CT scans are collected, used for assessing mispositioning, disruption, and loosening of the screws after spine surgery.} 
by collecting CT images from real clinical practice, which covers a wide range of diseased cases. The dataset will serve as a new benchmark to evaluate the effectiveness and performance of CT reconstruction from biplanar X-rays.

% \item We introduce a new regression model based on the implicit conditioning mechanism and conduct extensive experiments on three publicly available datasets and the LumbarV dataset. \textit{DiffuX2CT} outperforms the priors by a large margin and exhibits exciting performance in reconstructing realistic and faithful CT images from biplanar X-rays.
\end{itemize}

\section{Related Work}
\label{sec:related}
\textbf{CT Reconstruction from X-rays.}
Classical CT reconstruction methods, such as filtered back projection and iterative reconstruction, require a large number of X-ray projections taken from all viewing angles around the body to generate a high-quality volume \cite{herman2009fundamentals} but encounter severe distortions under ill-posed settings. Subsequent model-based iterative reconstruction (MBIR) methods \cite{venkatakrishnan2013model,mohan2015timbir,huang2018scale,schofield2020image,venkatakrishnan2021algorithm} are dedicated to using less number of X-rays and maintaining high sample quality, but they still can not address the unmet demand for real-time imaging while substantially reducing radiation. The success of deep learning methods makes it possible to reconstruct CT images using ultra-sparse X-rays. Regression-based methods \cite{henzler2018single,shen2019patient,wu2021reconstruction,shiode20212d,kasten2020end,ge2022x,zhang2023xtransct} directly learn a mapping from X-rays to CT through specialized neural networks with a mean squared error (MSE) loss. \cite{henzler2018single,shen2019patient,wu2021reconstruction} explore the feasibility of using a single X-ray projection to predict a 3D volume representation.
X-CTRSNet \cite{ge2022x} and XTransCT \cite{kasten2020end} achieve higher accuracy and show that using biplanar X-ray inputs can be advantageous for CT reconstruction. However, these regression-based methods perform poorly in generating high-quality CT images with fine details.
Recently, GAN-based methods \cite{ying2019x2ct,jiang2021reconstruction} have been proposed to improve the sample quality of generated CT images. For instance, X2CT-GAN \cite{ying2019x2ct}
increases the data dimension of 2D X-rays to 3D CT and introduces adversarial loss for realistic CT reconstruction. 
Despite the ability to generate relatively fine details, these GAN-based methods still suffer from severe artifacts and struggle to capture complex data distributions, yielding unnatural textures. 

\textbf{3D Generative Models.}
Generative Adversarial Networks (GANs) \cite{goodfellow2014generative}, Variational Auto-Encoders (VAEs) \cite{kingma2013auto}, and DMs have been widely used to model the explicit 3D data distribution, such as voxel grids \cite{brock2016generative,wu2016learning}, point clouds \cite{cai2020learning,xie2021style,luo2021diffusion}, and meshes \cite{liao2020towards,liu2023meshdiffusion}. The recent advancement of Neural Radiance Fields (NeRF) \cite{mildenhall2021nerf, barron2021mip} further facilitates the emergence of numerous works on building 3D structures through learning implicit representations \cite{li2023diffusion,chen2019learning,park2019deepsdf,sitzmann2020implicit,zeng2022fnevr,chan2021pi,deng2022gram,schwarz2020graf,zhang2024global}. In particular, EG3D \cite{chan2022efficient} first proposes the concept of hybrid explicit–implicit tri-plane representation for efficient 3D-aware generation from 2D data. Subsequently, Rodin \cite{wang2023rodin} and PVDM \cite{yu2023video} generate tri-plane representations for synthesizing 3D digital avatars and videos, respectively. In comparison, our objective is to extract the tri-plane representations from biplanar orthogonal X-rays for capturing the 3D structural information.

\textbf{Diffusion Probabilistic Models.}
Deep diffusion models (DMs) are first introduced by Sohl-Dickstein et al. \cite{sohl2015deep} as a novel generative model that generates samples by gradually denoising images corrupted by Gaussian noise. Recent DM advances have demonstrated their superior performance in image synthesis, including DDPM \cite{ho2020denoising}, DDIM \cite{song2020denoising}, ADM \cite{dhariwal2021diffusion}, LSGM \cite{vahdat2021score}, LDM \cite{rombach2022high}, and DiT \cite{peebles2022scalable}.
DMs also have achieved state-of-the-art performance in other synthesis tasks, such as text-to-image generation in GLIDE \cite{nichol2021glide}, DALLE-2 \cite{ramesh2022hierarchical}, and Imagen \cite{saharia2022photorealistic}, speech synthesis \cite{kong2020diffwave, liu2022diffsinger}, and super-resolution in SR3 \cite{saharia2022image} and IDM \cite{gao2023implicit}. Moreover, DMs have been applied to text-to-3D synthesis in DreamFusion \cite{poole2022dreamfusion} and its follows \cite{lin2022magic3d,wang2023score,tang2023make,wang2023prolificdreamer,li2024zone,zhu2023hifa,zeng2023ipdreamer}, video synthesis \cite{ho2022video,ho2022imagen,yu2023video,blattmann2023align,wu2023tune}, text-to-motion generation \cite{tevet2022human}, and other 3D object synthesis in RenderDiffusion \cite{anciukevivcius2022renderdiffusion}, diffusion SDF \cite{li2023diffusion}, and 3D point cloud generation \cite{luo2021diffusion}.
Recently, several works \cite{song2021solving,chung2022improving,chung2023solving,zeng2024controllable,lee2023improving,liu2023dolce} explore their potential in solving inverse problems in medical scenarios, such as sparse-view computed tomography (SV-CT), limited-angle computed tomography (LACT), and compressed-sensing MRI, \emph{etc.} We would like to highlight that our DiffuX2CT is fundamentally different from these diffusion-based CT reconstruction methods in that DiffuX2CT can extract faithful 3D conditions from 2D X-rays, while they still rely on the CT scanners to obtain the 3D conditions. Moreover,  their 2D denoising model's architectural design may not effectively handle the global coherence across the CT images.
\begin{figure*}[t]
\begin{center}
\includegraphics[width=0.95\textwidth]{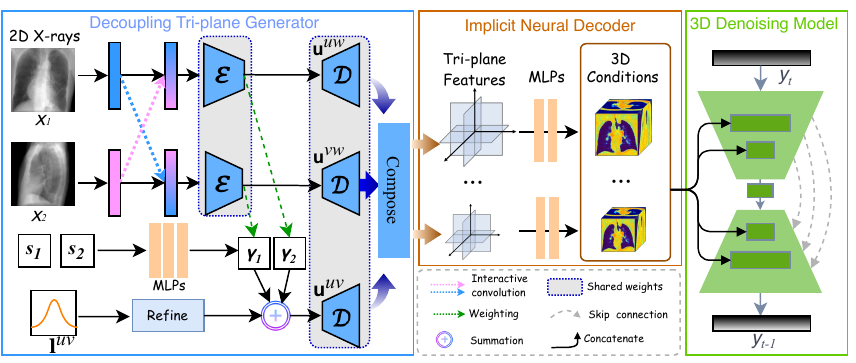}
   \caption{Overview of our \textit{DiffuX2CT}. Taking 2D biplanar orthogonal X-rays as inputs, the proposed implicit conditioning mechanism is implemented with a tri-plane decoupling generator and an implicit neural decoder to recover 3D structure information. By incorporating the 3D conditions, the 3D conditional denoising model iteratively generates CT images while preserving consistent structures provided in X-rays.}
\label{fig:overview}
\end{center}
\end{figure*}

\section{Methodology}
\label{sec:method}
This section presents the \textit{DiffuX2CT} approach, a 3D conditional diffusion model with an implicit conditioning mechanism (ICM), as shown in Fig.~\ref{fig:overview}. \textit{DiffuX2CT} is trained to recover faithful 3D structures from biplanar orthogonal X-rays (the front view and the side view) to generate high-quality 3D CT images. 
% The overall architecture of \textit{DiffuX2CT} is 

\subsection{Problem Statement}
\label{sec:overview}
Given a CT image $\mathbf{y}$ and paired biplanar X-rays of a front view $\mathbf{x}_1$ and a side view $\mathbf{x}_2$, \textit{DiffuX2CT} aims to learn a parametric approximation to the data distribution $p(\mathbf{y}|\mathbf{x}_1,\mathbf{x}_2)$ through a Markov chain of length $T$. Following \cite{saharia2022image}, we define the forward Markovian diffusion process $q$ by adding Gaussian noise as:
\begin{equation}
\begin{aligned}
% q\left(\mathbf{y}_{1: T} | \mathbf{y}_0\right) &=\prod_{t=1}^T q\left(\mathbf{y}_t | \mathbf{y}_{t-1}\right),  \\
q\left(\mathbf{y}_t | \mathbf{y}_{t-1}\right) &=\mathcal{N}\left(\mathbf{y}_t | \sqrt{1-\beta_t} \mathbf{y}_{t-1},\beta_t \mathbf{I}\right),
\end{aligned}
\end{equation}
where $\beta_t \in(0,1)$ are the variances of the Gaussian noise in $T$ iterations. 

In the training process, \textit{DiffuX2CT} is trained to infer the conditional distributions $p_\theta\left(\mathbf{y}_{t-1} | \mathbf{y}_t, \mathbf{x}_1,\mathbf{x}_2\right)$ to denoise the latent features sequentially. Formally, the inference process can be conducted as a reverse Markovian process from Gaussian noise $\mathbf{y}_T \sim \mathcal{N}(\mathbf{0}, \mathbf{I})$ to a target volume $\mathbf{y}_0$ as:
\begin{equation}
\begin{aligned}
% p_\theta\left(\mathbf{y}_{0: T} | \mathbf{x}_1,\mathbf{x}_2\right) &=p\left(\mathbf{y}_T\right) \prod_{t=1}^T p_\theta\left(\mathbf{y}_{t-1} | \mathbf{y}_t, \mathbf{x}_1,\mathbf{x}_2\right),\\
% p\left(\mathbf{y}_T\right) &=\mathcal{N}\left(\mathbf{y}_T | \mathbf{0}, \mathbf{I}\right), \\
p_\theta\left(\mathbf{y}_{t-1} | \mathbf{y}_t, \mathbf{x}_1,\mathbf{x}_2\right) &=\mathcal{N}\left(\mathbf{y}_{t-1} | \mu_\theta\left(\mathbf{x}_1,\mathbf{x}_2, \mathbf{y}_t, t\right), \sigma_t^2 \mathbf{I}\right).
\end{aligned}
\end{equation}

As shown in Fig.~\ref{fig:overview}, a 3D U-Net architecture is adopted as the denoising model that encodes the noisy image $\mathbf{y}_t$ into multi-resolution feature maps.
Meanwhile, ICM is designed to recover multi-resolution 3D structural features from 2D X-rays $\mathbf{x}_1,\mathbf{x}_2$ for conditioning the generation process of the denoising model.
Overall, \textit{DiffuX2CT} offers an elegant, end-to-end framework that unifies the iterative denoising process with 3D conditional learning from 2D inputs, without involving extra prior models.

\subsection{Implicit Conditioning Mechanism}
\label{sec:cond}
Extracting 3D structural information from 2D X-rays is crucial for reconstructing faithful CT images that have consistent details with the inputs. An intuitive idea is to first extract 2D features of X-rays, and then expand them to 3D conditions by duplicating along the channel dimension, and directly fuse them by pixel-wise addition, as done in X2CT-GAN \cite{ying2019x2ct}. This way, 3D structure information would not be well preserved. In contrast, we develop an implicit conditioning mechanism with a tri-plane decoupling generator and an implicit neural function.

\subsubsection{Tri-plane Decoupling Generator}
To implement our conditioning mechanism, we introduce a new formulation of tri-plane features for X-rays, which represents the 3D feature using three axis-aligned orthogonal planes, denoted by $\{\mathbf{u}^{uv}\in\mathbb{R}^{H \times W \times C},\mathbf{u}^{uw}\in\mathbb{R}^{H \times D \times C},\mathbf{u}^{vw}\in\mathbb{R}^{W \times D \times C}\}$, where $H, W, D$ denote the spatial resolution and $C$ is the number of channels. 
% \lxh{
Given that the input front view $\mathbf{x}_1$ and side view $\mathbf{x}_2$ can be directly utilized for extracting the $uw$ and $vw$ feature planes $\mathbf{x}^{uw}$ and $\mathbf{x}^{vw}$, we propose to generate the tri-plane features in a decoupling manner. Furthermore, to capture the spatial correlation between $\mathbf{x}^{uw}$ and $\mathbf{x}^{vw}$, we design an interleaved convolution to mutually enhance $\mathbf{x}_1$ and $\mathbf{x}_2$. To model the third feature plane $\mathbf{x}^{uv}$, we introduce two view modulators for refining and combining $\mathbf{x}^{uw}$ and $\mathbf{x}^{vw}$. The combination weighted by view modulators inherently guarantees the spatial correlation between $\mathbf{x}^{uv}$ and the other two feature planes. However, reconstructing the absent $\mathbf{x}^{uv}$ solely through these two feature planes would not be sufficient. Hence we include a learnable embedding with a Gaussian perturbation to collect statistical information on the $uv$ plane. Lastly, a lightweight decoder is employed to decode  $\mathbf{x}^{uv}$, $\mathbf{x}^{uw}$, and $\mathbf{x}^{vw}$, consequently producing $\mathbf{u}^{uv},\mathbf{u}^{uw}$, and $\mathbf{u}^{vw}$.
% }

% first leverage an interleaved convolution to mutually enhance $\mathbf{x}_1$ and $\mathbf{x}_2$, and then extract their tri-plane features separately. Meanwhile, unlike the simple concatenating operation, we learn two corresponding view modulators to modify and fuse the tri-plane features.

\textbf{Interleaved Convolution.} Since both $\mathbf{x}_1$ and $\mathbf{x}_2$ contain the information of the $w$ axis, we propose to capture their spatial correlated features by incorporating each other's information before inputting them into the implicit encoder.
Following a stem layer instantiated to extract initial features, we conduct the axis-wise pooling on $\mathbf{x}_1$ and $\mathbf{x}_2$ to obtain the vectors $\mathbf{x}_1^{w}, \mathbf{x}_2^{w} \in \mathbb{R}^{1 \times D \times C}$ along the $w$ axis. They are subsequently expanded to the original 2D dimension by replicating the vectors along row dimension, deriving the auxiliary features $\mathbf{x}_1^{(\cdot)w}\in \mathbb{R}^{W \times D \times C}$, $\mathbf{x}_2^{(\cdot)w} \in \mathbb{R}^{H \times D \times C}$. 
Then the interleaved convolution and the plane feature encoder are successively applied to process the enhanced features $\mathbf{\hat{x}}_1=\operatorname{Concat}(\mathbf{x}_1, \mathbf{x}_2^{(\cdot)w})$ and $\mathbf{\hat{x}}_2=\operatorname{Concat}(\mathbf{x}_2, \mathbf{x}_1^{(\cdot)w})$ for extracting $\mathbf{x}^{uw}$ and $\mathbf{x}^{vw}$.
% Note that the two feature planes are parallel extracted with the same plane feature encoder.

\textbf{View Modulators and Learnable Embedding.} 
% \lxh{
To obtain the absent $uv$ feature plane, 
% Considering the different focuses of $\mathbf{u}_1^{(i)}$ and $\mathbf{u}_2^{(i)}$ (\emph{i.e.} the tri-plane features from $\mathbf{x}_1$ contain richer $uw$ plane information), directly concatenating them can be insufficient. To address this issue, 
we initialize two view embeddings with $s_1=(1,0,1)$ for $\mathbf{x}_1$ and $s_2=(0,1,1)$ for $\mathbf{x}_2$, and map them to two view modulators $\gamma_{1},\gamma_{2},$ with the adaptive multi-layer perceptrons (MLPs). Subsequently, $\gamma_{1}$ and $\gamma_{2}$ are normalized by the L2 norm, and then used to modulate  $\mathbf{x}^{uw}$ and  $\mathbf{x}^{vw}$ channel-wisely and fuse them as the raw $uv$ feature plane  $\mathbf{\bar{x}}^{uv}$. Formally, the modulation process can be summarized as follows:
% }
% \begin{equation}
% \begin{aligned}
% {\mathbf{\gamma }} = Reshape(\operatorname{MLP} (s_1,s_2)), \\
% \end{aligned}
% \end{equation}
% \begin{center}
\begin{align}
    {\bar{\gamma}_{1}} = \frac{\left|\gamma_{1}\right|}{\sqrt{\gamma_1^{(2}+\gamma_2^{2}+\delta}}, 
    {\bar{\gamma}_{2}} = \frac{\left|\gamma_{2}\right|}{\sqrt{\gamma_1^{2}+\gamma_2^{2}+\delta}},
\end{align}
% \end{center}
\begin{equation}
\begin{aligned}
    {\mathbf{\bar{x}}^{uv}} =\bar{\gamma}_{{1}} \cdot \mathbf{x}^{uw}+\bar{\gamma}_{{2}} \cdot \mathbf{x}^{vw},
\end{aligned}
\end{equation}
% \lxh{
where $\delta=1 e-8$ is introduced to avoid zero denominators. Furthermore, considering the information loss in $uv$ plane, we adopt a sine distribution encoding $\mathbf{sine}$ and inject a Gaussian noise disturbance $\mathbf{z}$ to construct a learnable embedding $\mathbf{l}^{uv}$, which can supply additional information for generating the final $uv$ feature plane $\mathbf{u}^{uv}$. As a result, $\mathbf{{u}}^{uv}$ can be obtained by:
\begin{equation}
\begin{aligned}
    \mathbf{{x}}^{uv} = \mathbf{\bar{x}}^{uv} + \operatorname{CONV} (\mathbf{sine} + \mathbf{z}),
\end{aligned}
\end{equation}
% \lxh{
where $\operatorname{CONV}$ is a convolution layer with activation to refine the initial learnable embedding. Finally, we employ three simple decoders with shared weights, concurrently decoding $\mathbf{{x}}^{uv}$, $\mathbf{{x}}^{uw}$, and $\mathbf{{x}}^{vw}$ to
generate the multi-resolution tri-plane features $\mathbf{u}^{(i)}=\{\mathbf{u}^{uv,{(i)}},\mathbf{u}^{uw,{(i)}},\mathbf{u}^{vw,{(i)}}\}$, where $i \in\{1, \cdots, N\}$ denotes the index with different outputs, and $N$ is the number of stage depths in the decoders.
% }
\subsubsection{Implicit Neural Decoder}
Based on the expressive tri-plane features $\mathbf{u}^{(i)}$ with rich explicit 3D information, we aim at recovering the 3D structural information used as the condition for the CT generation. To accomplish this, we leverage the appealing property of implicit neural function by encoding the tri-plane features as a query function based on the coordinates to acquire the 3D structural representation. 
Specifically, we assign $\mathbf{u}^{(i)}$ with the assumed 3D spatial coordinates $\mathbf{c}^{\left(i\right)} \in \mathbb{R}^{(D \times H \times W) \times 3}$ as a reference, where $D \times H \times W$ denotes the number of 3D points.
Given the coordinates of each 3D point, we sample three corresponding feature vectors with bilinear interpolation by projecting them onto each plane and aggregating the retrieved vectors via summation. Afterward, the final 3D structural representation $\mathbf{h}^{(i)}\in \mathbb{R}^{D \times H \times W \times C}$ can be derived using a two-layer lightweight MLP decoder by:
\begin{equation}
\begin{aligned}
    {\mathbf{h}^{(i)}} = Reshape\{\operatorname{MLP}[\mathbf{u}^{(i)}(\mathbf{c}^{(i)})]\},
\end{aligned}
\end{equation}
where $\mathbf{u}^{(i)}(\mathbf{c}^{(i)})$ represents the aggregated features of the three vectors projected onto each plane. Note that we do not input the spatial coordinates into the lightweight MLP decoder.
As a result, the multi-resolution 3D representations $\mathbf{h}^{(i)}$ can serve as faithful 3D conditions for guiding the CT generation.

\subsection{Model Architecture}
\label{sec:architecture}
Given the feature maps $\mathbf{f}_{\text{down}}^{\left(i\right)}$ and $\mathbf{f}_{\text{up}}^{\left(i\right)}$ from the encoder and decoder of the 3D U-Net denoising model, where $i \in\{1, \cdots, N\}$  and $N$ is the number of stage depths in the 3D U-Net, unlike the cross-attention mechanism that introduces additional computational cost, we directly concatenate them with the 3D conditions $\mathbf{h}^{(i)}$, respectively, so that CT features are synchronously generated according to the structural information in the conditions. While the 3D denoising model possesses more weight parameters compared to the 2D U-Net, it offers superior global coherence throughout the 3D volume. 
Moreover, the 2D U-Net synthesizes each slice in CT one by one for reconstructing a CT image, leading to a much slower inference time than our 3D model.
To reduce the computation cost, we decrease the number of U-Ne layers by setting $N$ to 3 and employ the shifted-window attention layer \cite{liu2022video} rather than self-attention in the last stage of the 3D U-Net.

% In comparison with DiffuX2CT, though ICM-\textit{Reg} does not require the 3D encoder and generates a CT volume at once without an iterative denoising process, 
% it yields lower-quality samples lacking high-frequency details.
% it does not circumvent the limitations of regression techniques that often yield lower-quality samples lacking high-frequency details.

\subsection{Training Objective}
\label{sec:optimization}
\textbf{Noise Prediction Loss.}
By introducing $\mathbf{x}_1, \mathbf{x}_2$ as the conditions and their corresponding view embeddings $s_1,s_2$ in the reverse diffusion process, we optimize the denoising model as a noise estimator $\epsilon_\theta$ to reconstruct the target CT image $\hat{\mathbf{y}}_0$ from a pure noise. The noise prediction loss is defined as:
\begin{equation}
\mathcal{L}_{simple} = E_{(\mathbf{x}_1,\mathbf{x}_2,\mathbf{y})}E_{\epsilon,t}\left[\left\|\epsilon-\epsilon_\theta\left(\mathbf{y}_t, t, \mathbf{x}_1, \mathbf{x}_2,s_1,s_2\right)\right\|_1^1\right],
\end{equation}
where $\epsilon \sim \mathcal{N}(0,1), t\sim [1, T]$, and $(\mathbf{x}_1,\mathbf{x}_2,\mathbf{y})$ are paired X-rays and CT. 

\textbf{Geometry Projection Loss.} Furthermore, we draw inspiration from \cite{ying2019x2ct} and introduce a geometry projection loss $\mathcal{L}_{proj}$. Specifically, given the predicted noise $\hat{\epsilon}$ and the timestep $t$, we can derive the $\hat{\mathbf{y}}_0$ by:
\begin{equation}
\hat{\mathbf{y}}_0=\left(\mathbf{y}_t-\left(1-\alpha_t\right) \cdot \hat{\varepsilon}\right) / \sqrt{\alpha_t},
\end{equation}
where $\alpha_t=\prod_{i=1}^t \left( 1 - \beta_i\right)$. To streamline the process and focus on general geometry consistency, we employ three orthogonal projections to calculate the loss:
\begin{equation}
\begin{aligned}
\mathcal{L}_{proj} &= \frac{1}{3}(||\mathcal{P}_A(\mathbf{y}_0)-\mathcal{P}_A(\hat{\mathbf{y}}_0)||_1^1 + ||\mathcal{P}_C(\mathbf{y}_0)-\mathcal{P}_C(\hat{\mathbf{y}}_0)||_1^1
+ ||\mathcal{P}_S(\mathbf{y}_0)-\mathcal{P}_S(\hat{\mathbf{y}}_0)||_1^1).
\end{aligned}
\end{equation}
Putting everything together, we obtain the training objective as follows:
\begin{equation}
\mathcal{L}_{total} = \mathcal{L}_{simple} + \lambda \mathcal{L}_{proj},
\end{equation}
where we set $\lambda=1/T$ in our experiments.

\begin{figure*}[t]
\centering
\includegraphics[width=.99\textwidth]{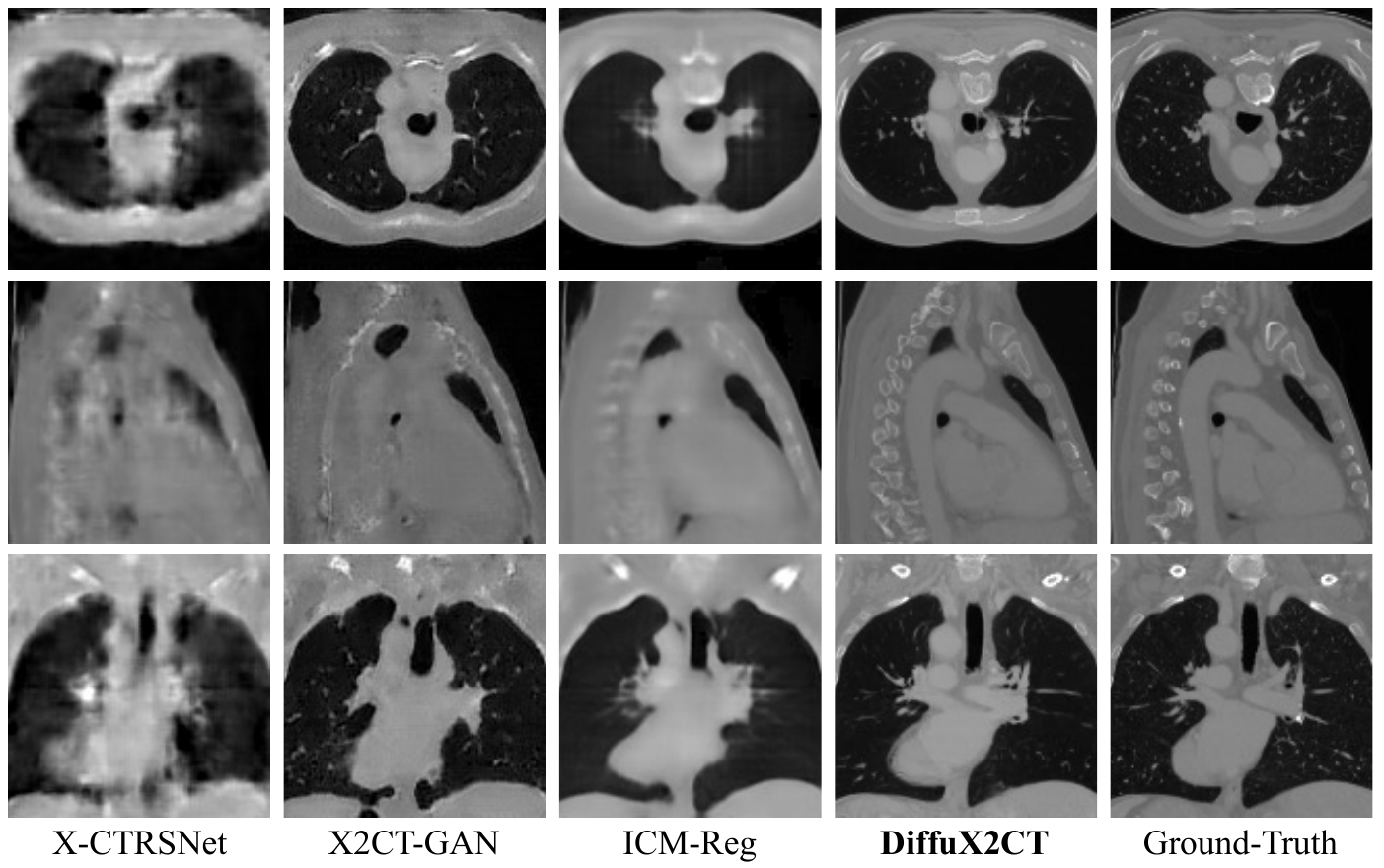} % Reduce the figure size so that it is slightly narrower than the column.
\caption{Qulitative comparisons on the CTSpine1K dataset.
% The black part is the contrast agent. 
From top to bottom, we show the axial, sagittal, and coronal views. Best viewed by zoom in.}
\label{fig:spine}
\end{figure*}

\section{Experiments}
\label{sec:experiments}
In addition to the extensive experiments described in this section, we also provide more implementation details and results in the supplementary materials.

\subsection{Datasets}
\label{sec:dataset}
We conduct extensive experiments on four datasets, including a newly collected lumbar dataset and three publicly available datasets.
% encompassing the practical applications of implant localization and visualization of the chest, spine, and pelvis.

\textbf{Newly Collected Lumbar Dataset.}
In our experiments, we collect a new lumbar vertebra dataset, called LumbarV, which comprises 268 3D CT images of different patients. Each patient has implants inserted in the vertebrae. 
We randomly select 231 CT images for training and leave the remaining 37 CT images for test. To ensure consistency, we initially resampled all CT images to an isotropic voxel resolution of $2\times 2 \times 2 ~ mm^3$ and then center-cropped them to a fixed size of $128 \times 128 \times 128$.
Then, we adopt CT value clipping to limit the CT value range to [-1024, 1500], and min-max normalization to scale data to the [-1, 1] range. 
To construct the paired data of CT and biplanar X-ray data for training and testing, we follow previous works \cite{shen2019patient,ying2019x2ct} and synthesize the corresponding biplanar X-rays from imagetric CT images using digitally reconstructed radiographs (DRR) \cite{13ddr}.

\begin{figure*}[t]
\centering
\includegraphics[width=.99\textwidth]{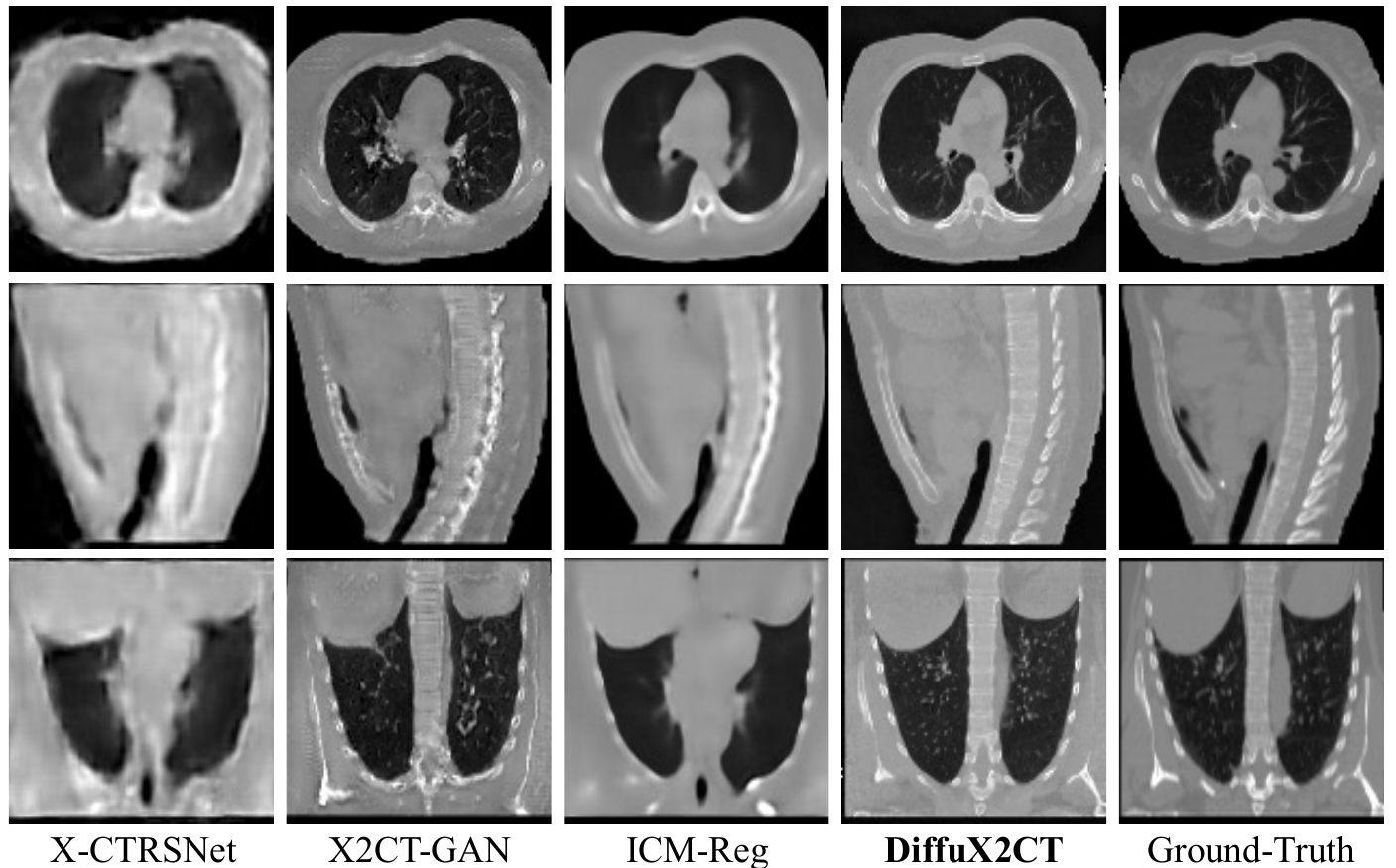} % Reduce the figure size so that it is slightly narrower than the column.
\caption{Qulitative comparisons on the LIDC-IDRI dataset.
% The black part is the contrast agent. 
From top to bottom, we show the axial, sagittal, and coronal views. Best viewed by zoom in.}
\label{fig:chest}
\end{figure*}

\begin{figure*}[t]
\centering
\includegraphics[width=.99\textwidth]{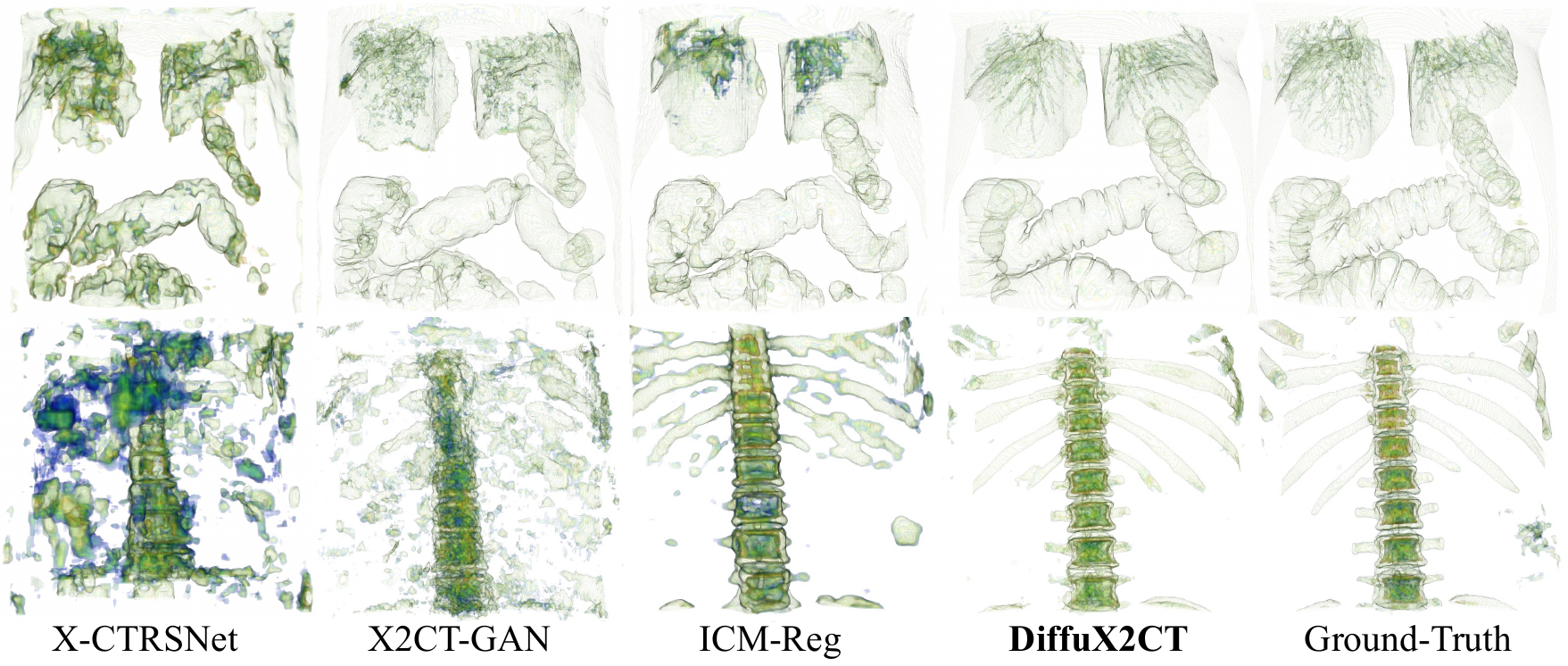} % Reduce the figure size so that it is slightly narrower than the column.
\caption{Rendering comparisons on the CTSpine1K datasets. The first and second row show the visualizations of soft tissues and bones, respectively.}
\label{fig:render}
\end{figure*}

\begin{table*}[t]
\centering
\begin{tabular}{clcccccc}
\toprule
Dataset                   & Method & PSNR$\uparrow$ & PSNR3D$\uparrow$ & SSIM$\uparrow$ & SSIM3D$\uparrow$ & FID $\downarrow$& LPIPS  $\downarrow$ \\ \midrule
\multirow{3}{*}{LumbarV}  & X-CTRSNet \cite{ge2022x}&      31.67&  27.98&      0.7738&    0.7502&   75.28  &  0.2957  \\
                          & X2CT-GAN \cite{ying2019x2ct}  &  34.27   &   27.19     &   0.7456   &     0.6518  &   35.04  &  0.1554   \\ 
                          & \cellcolor{softblue}\textit{\textit{DiffuX2CT}} & \cellcolor{softblue}  \textbf{34.32}   &  \cellcolor{softblue}   \textbf{28.01 } & \cellcolor{softblue} \textbf{0.7984 }   & \cellcolor{softblue}  \textbf{ 0.8544   } & \cellcolor{softblue}   \textbf{13.23 }&  \cellcolor{softblue}  \textbf{0.1198  }    \\ \midrule
\multirow{3}{*}{CTPelvic1K} & X-CTRSNet \cite{ge2022x}&      23.89&        23.27&      0.6512&        0.6103&  91.64  &    0.5065     \\
                          & X2CT-GAN \cite{ying2019x2ct}  &  23.33    &   22.64     &   0.5922   &   0.5571     &    28.88   &    0.2524     \\ 
                          & \cellcolor{softblue}\textit{\textit{DiffuX2CT}} & \cellcolor{softblue} \textbf{ 23.91 }   &  \cellcolor{softblue}  \textbf{23.27 }  &\cellcolor{softblue}   \textbf{0.6584  } & \cellcolor{softblue} \textbf{   0.7650 }   & \cellcolor{softblue}  \textbf{ 7.78 }  &  \cellcolor{softblue}    \textbf{ 0.1696 }  \\ 
                          \midrule
\multirow{3}{*}{CTSpine1K}  & X-CTRSNet \cite{ge2022x}&    21.34&        20.84&      0.5355&       0.5159 & 255.98  &      0.3981   \\
                          & X2CT-GAN \cite{ying2019x2ct}  &   20.60   &   20.16     &   0.4841   &     0.5052   &   50.63  &   0.2221  \\ 
                          & \cellcolor{softblue}\textit{\textit{\textit{DiffuX2CT}}}  & \cellcolor{softblue}  \textbf{ 21.53 }  &  \cellcolor{softblue}    \textbf{21.12 } & \cellcolor{softblue} \textbf{0.5924  }  & \cellcolor{softblue}  \textbf{ 0.7099 }   & \cellcolor{softblue}  \textbf{  8.90 }&  \cellcolor{softblue}    \textbf{ 0.1673 }   \\ \midrule
\multirow{3}{*}{LIDC-IDRI} & X-CTRSNet \cite{ge2022x}&  23.95&   \textbf{ 22.35}& 0.6223&   0.6228&  143.73  &     0.2784 \\
                          & X2CT-GAN \cite{ying2019x2ct}  &  26.02    &   21.17     &  0.6086    &   0.6199     &   16.65   &     0.1354     \\ 
                          & \cellcolor{softblue}\textit{\textit{\textit{DiffuX2CT}}} & \cellcolor{softblue}   \textbf{26.35}    &  \cellcolor{softblue} {21.15}   &\cellcolor{softblue}   \textbf{0.6872}   & \cellcolor{softblue}    \textbf{0.7423}    & \cellcolor{softblue}   \textbf{4.52}  &  \cellcolor{softblue}     \textbf{0.1217}   \\
                          \bottomrule
\end{tabular}
\caption{Quantitative comparisons on the LumbarV and three public datasets. }
\label{tab:quantitative}
\end{table*}

\textbf{Public Datasets.}
To make a comprehensive evaluation, we also perform experiments on three publicly available datasets: (1) \textbf{LIDC-IDRI dataset} \cite{armato2011lung} contains 1,018 chest CT images. We follow the work of X2CT-GAN \cite{ying2019x2ct} to split 917 CT images for training and 101 for test; (2) \textbf{CTSpine1K \cite{deng2021ctspine1k}} is collected from four open sources, including COLONOG \cite{johnson2008accuracy}, HNSCC-3DCT-RT \cite{nolan2022head}, MSD Liver \cite{simpson2019large}, and COVID-19 \cite{harmon2020artificial}. We randomly select 912 images for training and the remaining 102 for testing. 
(3) \textbf{CTPelvic1K \cite{liu2021deep}} is constructed for pelvic bone segmentation with annotation labels for the lumbar spine, sacrum, left hip, and right hip. We delete the duplicate data with CTSpine1K and randomly split 328 of the remaining data as the training set and 50 as the test set.

\subsection{Implementation Details} 
\textbf{Compared Methods.}
We compare our \textit{\textit{DiffuX2CT}} quantitatively and qualitatively with a regression-based method X-CTRSNet \cite{ge2022x} and a GAN-based method X2CT-GAN \cite{ying2019x2ct}. For X2CT-GAN, we use the official code provided by the authors to perform experiments. For X-CTRSNet, we reproduce the model code according to the paper, due to the absence of the official code.
Particularly, since the proposed ICM formulates an effective way to recover 3D information from the 2D X-rays, we can easily develop a regression model, termed ICM-\textit{Reg}, by combining ICM with a simple 3d decoder. Detailed model architecture is shown in the supplementary.

\textbf{Training Details.}
We train our \textit{\textit{DiffuX2CT}} in an end-to-end manner. 
Following the vanilla DDPM \cite{ho2020denoising}, we use the Adam optimizer with a fixed learning rate of $2e-4$. The number of training iterations is set to $200,000$ for the LumbarV and CTPelvic1K datasets and $800,000$ for the LIDC-IDRI and CTSpine1K datasets.
For training ICM-\textit{Reg}, we use the same optimizer to train $200$ epochs.
We utilize a dropout rate of 0.2 and two 80GB NVIDIA RTX A800 GPUs for all experiments.

\textbf{Evaluation Metrics.}
The evaluation metrics include two distortion-based metrics PSNR, SSIM, and two perceptual metrics FID and LPIPS. We calculate PSNR and SSIM in two ways: (1) directly computing the values of 3D data, denoted as PSNR3D and SSIM3D; and (2) averaging the values of all 2D slices along the three axes. It is worth noting that we use the image encoder of MedCLIP \cite{wang2022medclip} trained on the medical dataset, i.e., MIMIC-CXR \cite{johnson2019mimic}, which is appropriate for medical data to compute FID. 

\subsection{Comparisons with Prior Arts} 
We compare with two representative families of reconstruction models, i.e., generative and regression models. The performance of CT reconstruction methods is measured both qualitatively and quantitatively on the four datasets.

\textbf{Qualitative Results.}
Fig. \ref{fig:spine}, Fig. \ref{fig:chest}, and Fig. \ref{fig:render} show the qualitative comparisons with prior arts on the LIDC-IDRI and CTSpine1K datasets. The results of the regression model X-CTRSNet exhibit inferior sample quality.
Although the generative model, X2CT-GAN, can generate sharp textures, it often creates severe artifacts (\emph{e.g.,} incorrect heart shape and lack of aorta and pulmonary artery in Fig.~\ref{fig:chest}) and distortions (\emph{i.e.} unnatural bone structures in Fig. \ref{fig:render}).
Our regression version, ICM-\textit{Reg}, is effective in recovering accurate structures but suffers from over-smoothing textures and lacks high-frequency details.
In comparison, our \textit{DiffuX2CT} is capable of reconstructing clear CT textures and consistent structure with the ground-truth, far superior to the counterpart methods.

\textbf{Quantitative Results.}
Table~\ref{tab:quantitative} shows the quantitative comparisons with prior arts on the four datasets.
% The regression model ICM-\textit{Reg} exhibits the overall best performance in 。distortion-based metrics, while 
\textit{DiffuX2CT} exhibits the overall best results in terms of distortion-based metrics. This highlights the effectiveness of our implicit conditioning mechanism in extracting faithful structural information from 2D X-rays.
% \lxh{
Particularly, \textit{DiffuX2CT} performs much better than other methods in terms of SSIM3D, showcasing its strong capacity to reconstruct CT images that are globally similar to the ground-truth in 3D space.
% }
Moreover, \textit{\textit{DiffuX2CT}} surpasses others by a large margin in all perceptual metrics, demonstrating its superiority in reconstructing high-quality CT images. Although the regression-based models, e.g., X-CTRSNet, can deliver promising PSNR since they are typically optimized by directly minimizing the L1 or L2 losses, they often encounter subpar sample quality and deliver lower perception metric scores.

\subsection{Further analysis by 3D Segmentation} 
\begin{figure*}[t]
\centering
\includegraphics[width=\textwidth]{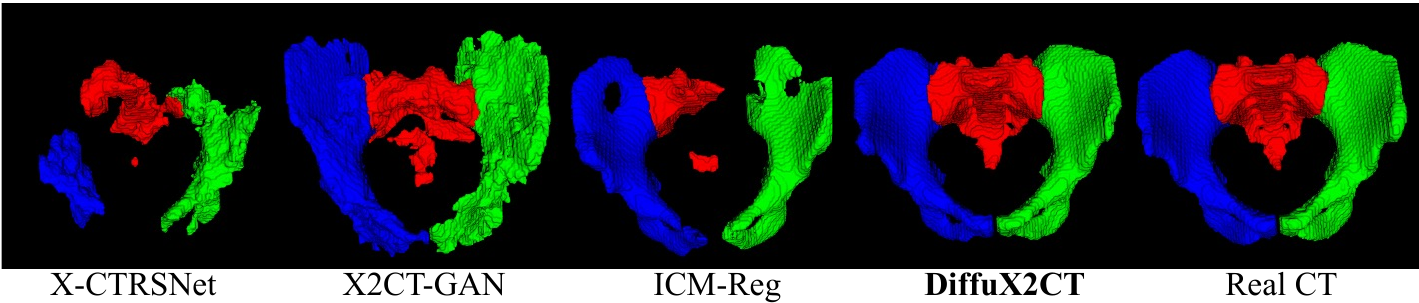} 
\caption{Comparisons by rendering the 3D segmentation masks of the reconstructed CT images. The illustrated images are selected from the CTPelvic1K dataset. Our \textit{DiffuX2CT} performs the best among compared methods.}
\label{fig:seg}
\end{figure*}

To intuitively demonstrate the effectiveness and practical significance of our \textit{\textit{DiffuX2CT}}, we evaluate the performance in 3D segmentation of the reconstructed CT images on the CTPelvic1K dataset. Specifically, we utilize a well-trained VNet \cite{abdollahi2020vnet} for 3D pelvic segmentation to segment the sacrum, coccyx, left and right pelvis in the CT images.
% ground-truth and reconstructed results.
The visualizations of inputs with segmentation masks are shown in Fig. \ref{fig:seg}. We can clearly observe that \textit{\textit{DiffuX2CT}} significantly surpasses other methods, capably reconstructing bone structures that align with real CT.
Moreover, we use the Dice Similarity Coefficient (DSC) as the evaluation metric for segmentation. Quantitative results of the three objectives are given in Table \ref{tab:seg}. \textit{\textit{DiffuX2CT}} outperforms the previous works by a large margin in all objectives, demonstrating its effectiveness in reconstructing CT images with commendably accurate shapes and positions from biplanar X-rays.

\begin{table}[t]
\begin{minipage}[t]{0.48\textwidth}
\centering
\makeatletter
\def\@captype{table}\makeatother
\setlength{\tabcolsep}{0.1mm}{
\scalebox{0.95}{
\begin{tabular}{lcccc}
\toprule
Data Source                & {Avg} & {Sac$\&$Coc} & {L.Pel}& {R.Pel} \\ \midrule
Real (oracle)                  & 90.19     & 87.55              & 91.62        &    91.41      \\ \midrule
X-CTRSNet    & 20.51     & 25.66              & 24.71        & 11.17   \\
X2CT-GAN  & 45.95   & 43.08              &  50.31       & 44.44                               \\ 
% \hline
\cellcolor{softblue}ICM-\textit{Reg}                & \cellcolor{softblue}47.59     & \cellcolor{softblue}31.98              &\cellcolor{softblue}56.82         & \cellcolor{softblue}53.97      \\ 
\cellcolor{softblue}\textit{\textit{DiffuX2CT}}       & \cellcolor{softblue}\textbf{74.10} &\cellcolor{softblue} \textbf{73.63}  & \cellcolor{softblue}\textbf{73.31}   &\cellcolor{softblue}\textbf{75.35} \\ 
\bottomrule
\end{tabular}}}
\caption{Quantitative results in terms of DSC for 3D segmentation. Higher DSC indicates better performance.}
\label{tab:seg}
% \end{table}
\end{minipage}
\hspace{1ex}
\begin{minipage}[t]{0.48\textwidth}
\centering
\makeatletter\def\@captype{table}\makeatother
% \begin{table}[h]
% \centering
\setlength{\tabcolsep}{1mm}{
\scalebox{0.95}{
\begin{tabular}{lccc}
\toprule
Method    & PSNR & SSIM & FID \\\midrule
Baseline  &  29.37  &  0.7034  &   17.15     \\
$w/o$ $\mathcal{L}_{proj}$ &33.62   &    0.7630  &  16.22     \\
$w/o$ V.M. & 32.09   &    0.7674   &   16.52   \\
$w/o$ $\mathbf{l}^{uv}$ &34.19   &   0.7781  & 14.51     \\
ResNet & 33.07    &  0.7593 &  13.77     \\
\cellcolor{softblue}\textit{\textit{DiffuX2CT}} & \cellcolor{softblue}\textbf{34.32} & \cellcolor{softblue}\textbf{0.7984} & \cellcolor{softblue}\textbf{13.23}\\ 
\bottomrule
\end{tabular}}}
\caption{Results of ablation study. V.M. denotes the view modulators.}
\label{tab:ablation}
\end{minipage}
\end{table}

\subsection{Ablation Study} 
We conduct ablation studies on the LumbarV dataset to demonstrate the effectiveness of the proposed ICM and the geometry projection loss $\mathcal{L}_{proj}$. Specifically, we construct five comparison models: (1) the baseline model, which is a 3D conditional diffusion model with a simple conditioning mechanism that extracts 3D conditions by a 2D encoder with expanding and pixel-wise addition operations like X2CT-GAN.
(2) \textit{\textit{DiffuX2CT}} trained from scratch without $\mathcal{L}_{proj}$. 
(3)-(4) \textit{\textit{DiffuX2CT}} trained without the view modulators and learnable embedding $\mathbf{l}^{uv}$ respectively.
(5) using a ResNetBlock to fuse the $uw$ and $vw$ feature planes without the view modulators.
As shown in Table \ref{tab:ablation}, all the proposed components contribute to the overall performance. The three models all yield good performance in perceptual metrics, illustrating the effectiveness of the 3D diffusion model in reconstructing high-quality CT images.
Notably, the implicit conditioning model significantly improves the performance in terms of PSNR and SSIM, demonstrating its great effectiveness in capturing 3D structural information from 2D X-rays.

\subsection{Clinical Case Study} 
\begin{figure}[t]
\centering
\includegraphics[width=0.99\linewidth]{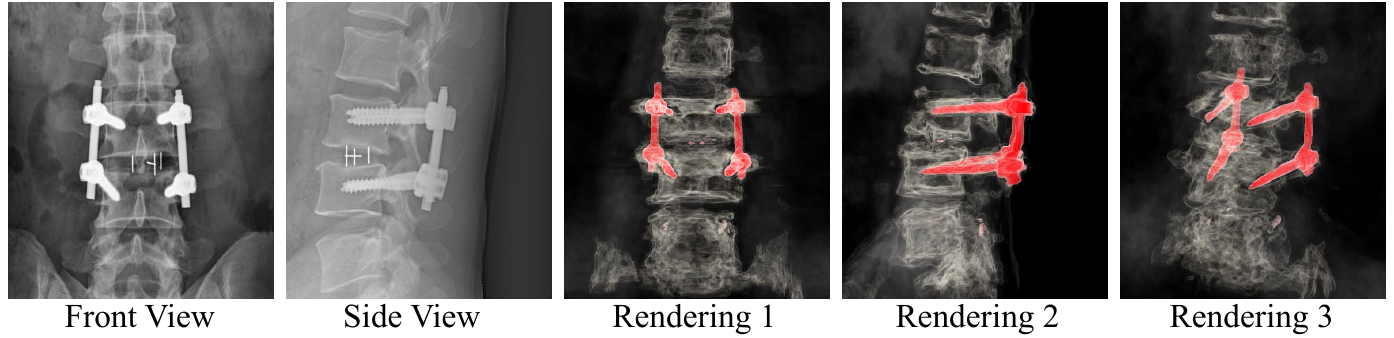} 
\caption{Visualization of the reconstructed CT from real X-rays. The first and second columns are two input X-rays collected from real clinical practice.}
\label{fig:case}
\end{figure}

% \lxh{
To demonstrate the clinical utility of the proposed  \textit{DiffuX2CT}, we explore its actual clinical application in lumbar vertebra surgery. In that case, orthogonal X-rays can be easily acquired with C-arm machines, but CT images are often unavailable. This poses a great challenge to accurately locate the position and shape of implants, thereby increasing the difficulty of surgery. 
Fortunately, \textit{DiffuX2CT} can potentially serve as a valuable alternative to reconstruct CT images.
To verify this, we assess the performance of our \textit{DiffuX2CT} on real-world lumbar X-ray data.
Due to the absence of corresponding ground truth for the X-rays, we limit the evaluation to qualitative analysis.
The reconstructed results of \textit{DiffuX2CT} are shown in Fig. \ref{fig:case}. 
% As the primary goal of this study is to generate a CT image from real X-rays, we adjust the contrast of the generated X-ray to simulate real-world data. 
Despite being trained on synthetic data, \textit{DiffuX2CT} successfully reconstructs a realistic lumbar CT image. Especially, it effectively recovers the structure of bones and locates the position of implants, which is well aligned with real-world biplanar X-rays. 
This case study demonstrates that \textit{DiffuX2CT} can potentially assist surgeons in observing the precise shapes and positions of bones and implants.

\section{Conclusion}
\label{sec:conclusion}
This paper presents a new 3D conditional diffusion model, DiffuX2CT, that achieves high-quality and faithful CT reconstruction from biplanar X-rays. It is a significant breakthrough in this area. Specifically, DiffuX2CT adopts a 3D denoising model to ensure a high sample quality and global coherence across the 3D volume. More importantly, DiffuX2CT adopts a new implicit conditioning mechanism, comprised of a tri-plane decoupling generator and an implicit neural decoder, that effectively captures 3D structure information from 2D X-rays. The 3D conditions enable DiffuX2CT to achieve faithful CT reconstruction.
Moreover, we construct a new lumbar dataset, named LumbarV, as a new benchmark for validating the significance and performance of CT reconstruction from biplanar X-rays. Extensive experiments on three public datasets and the LumbarV dataset demonstrate the effectiveness of our DiffuX2CT.

\textbf{Limitation.} 
The purpose of this work is not to replace the current CT reconstruction methods using hundreds and thousands of X-rays, but to offer a valuable alternative when performing CT scans is infeasible. Although DiffuX2CT can reconstruct high-quality CT images with faithful structures, \emph{i.e.}bones, implants, it may not always accurately synthesize soft tissues, such as pulmonary nodules. One potential solution to this problem would be to introduce more X-ray views to gather additional structural information on soft tissues.
\clearpage  % TODO FINAL: This \clearpage needs to be removed from both review and camera-ready versions.

\section*{Acknowledgements}
The work was supported by the National Key Research and Development Program of China (Grant No. 2023YFC3300029). This research was also supported by the Zhejiang Provincial Natural Science Foundation of China under Grant No. LD24F020007, Beijing Natural Science Foundation L223024, National Natural Science Foundation of China under Grant NO. 62076016, 62176068, and 12201024, “One Thousand Plan” projects in Jiangxi Province Jxsg2023102268, Beijing Municipal Science $\&$ Technology Commission, Administrative Commission of Zhongguancun Science Park Grant No.Z231100005923035. Taiyuan City "Double hundred Research action"  2024TYJB0127.
\bibliographystyle{splncs04}
\bibliography{main}
\end{document}

\title{\textit{DiffuX2CT}: Diffusion Learning to Reconstruct CT Images from Biplanar X-Rays} 

\titlerunning{DiffuX2CT}

\author{Xuhui Liu\inst{1} \and Zhi Qiao \inst{2} \and Runkun Liu \inst{2} \and Hong Li  \inst{1} \and Juan Zhang$^*$  \inst{1} \and \\
Xiantong Zhen$^*$ \inst{2} \and  Zhen Qian \inst{2} \and Baochang Zhang \inst{1,3} }

\authorrunning{X. Liu et al.}

\institute{Beihang University, Beijing, China \and
Central Research Institute, United Imaging Healthcare, Beijing, China\\  \and
Zhongguancun Laboratory, Beijing, China\\
\email{\{xhliu,zhang\_juan\}@buaa.edu.cn}, \email{zhenxt@gmail.com}}

\maketitle

\renewcommand{\thefootnote}{\fnsymbol{footnote}}
\footnotetext[1]{Corresponding authors.}

In this supplementary material, we first introduce the dataset preprocessing techniques in Section \ref{data} and additional implementation details in Section \ref{implementation}. Then, we provide the detailed model architecture of our regression model, ICM-\textit{Reg}, in Section \ref{tpnet}. Lastly, we present additional results on the four datasets in Section \ref{result}.
% Lastly, we show additional 3D segmentation results of the reconstructed CT images in Section \ref{seg}.
% and visualize additional reconstructed CT images from real X-rays in Section \ref{case}.
% The source code will be released soon.

\section{Dataset Establishment}
\label{data}
In our experiments, we need paired X-ray and CT data for training \textit{DiffuX2CT}. However, no such dataset is available and it is very challenging to collect the required dataset. Therefore,
we follow \cite{shen2019patient,ying2019x2ct} to introduce the digitally reconstructed radiograph (DRR) for synthesizing the biplanar X-rays from CT images. We elaborate on the DRR imaging technique and preprocessing details associated with the CTSpine1K \cite{deng2021ctspine1k} and CTPelvic1K \cite{liu2021deep} below.
\subsection{Digitally Reconstructed Radiographs (DRR)}
DRR simulates radiographic images by projecting 3D volume images onto a 2D image plane using a variety of ray tracing techniques. Current DRR algorithms mainly build upon ray casting, a classic volume rendering technique. It simulates the process of X-rays passing through the human body and being attenuated by the absorption of human tissues to generate DRR images.

The goal of our \textit{DiffuX2CT} is to reconstruct CT images from biplanar orthogonal X-rays. Accordingly, we need to simulate two DRR images of the front view and side view from the real CT image. For each imaging, DRR first assumes a source point as the conventional X-ray source, and then emits multiple virtual X-rays from this point source, passing through the 3D CT image and projecting onto a panel perpendicular to the center axis of the X-ray beam. The projection points on the panel are the accumulated results of all the intersections between the virtual X-rays and the CT image, corresponding to the pixels of the DRR image.

\subsection{Repurposing Existing Datasets}
\paragraph{CTSpine1K.}
CTSpine1K is a large-scale spine CT dataset, which consists of 1014 CT volumes from four open sources. We initially resample all CT images to an isotropic voxel resolution of $2 \times 2 \times 2$ $mm^{3}$ and then center-crop them to a fixed size of $128 \times 128 \times 128$. Then, we adopt CT value clipping to limit the CT value range to $[-1024, 1500]$, and min-max normalization to scale data to the $[-1, 1]$ range. Then we utilize DDR technologies to generate the front view and side view for each CT volume. To ensure data splitting (9:1) consistency, 912 cases are randomly sampled for training, and the remaining 102 cases are used for test.

\paragraph{CTPelvic1K.} CTPelvic1K is a large-scale pelvic CT dataset reorganized from 7 open source sub-datasets. We delete the COLONG sub-dataset that overlaps with CTSpine1K, and the remaining 365 CT volumes are used for our experiments. 
We preprocess all the CT images in the same manner as we do in CTSpine1K.
% As we preprocess in CTSpine1K, we initially resampled all CT images to an isotropic voxel resolution of $2 \times 2 \times 2$ $mm^{3}$ and then center-cropped them to a fixed size of $128 \times 128 \times 128$.  Then, we adopt CT value clipping to limit the CT value range to [-1024, 1500], and min-max normalization to scale data to the [-1, 1] range.   Then we adopted DDR technologies to generate anteroposterior view and lateral view for each CT volume. 
To ensure data splitting (9:1) consistency, 328 cases are randomly sampled for training, and the remaining 37 cases are used for test.

\section{Additional Implementation Details}
\label{implementation}
\subsection{Metrics}
\label{metric}
We provide a more detailed description of the evaluation metrics below:
\paragraph{Peak Signal-to-Noise Ratio (PSNR).}  PSNR reflects the image reconstruction quality between the maximum signal and the background noise. Since PSNR tends to penalize synthetic high-frequency, which are not well aligned with the ground-truth, the PSNR value (dB) has limitations in CT reconstruction evaluation.
The larger the PSNR value, the less image distortion.
\paragraph{Structure Similarity Index Measure (SSIM).}
From the perspective of image distortion modeling discussed in \cite{wang2004image}, SSIM implements the related theory of structural similarity by imitating the human visual system and is sensitive to the perception of local structural changes in the image. SSIM quantifies image properties from brightness, contrast, and structure, using mean to estimate brightness, variance to estimate contrast, and covariance to estimate structural similarity.
\paragraph{Learned Perceptual Image Patch Similarity (LPIPS).}
LPIPS aims to measure the perceptual similarity between the generated and real images by using their deep features. We utilize the pre-trained AlexNet \cite{krizhevsky2012imagenet} to extract the deep features for calculating the LPIPS values.
\paragraph{Frechet Inception Distance score (FID)} \cite{MartinHeusel2017GANsTB} evaluates the image quality by imitating human perception of image similarity. We utilize a pre-trained MedCLIP \cite{wang2022medclip} to compare the distributions of the generated images with those of the ground-truth images.

\subsection{Training Details of Compared Methods}
We use the official code of X2CT-GAN \cite{ying2019x2ct} while reproducing the model code for X-CTRSNet \cite{ge2022x} to perform experiments. Both methods are conducted in an NVIDIA A40. 
\paragraph{X2CT-GAN.}
Following its official codes, we use the Adam solver with momentum parameters $\beta_1=0.5$ and $\beta_2=0.999$. We train X2CT-GAN for a total of 300 epochs with a learning rate of 2e-4. Constrained by the GPU memory limit, the batch size is set to 4.
\paragraph{X-CTRSNet.} 
Due to the absence of the official codes, we reproduce the network code of X-CTRSNet according to its published paper. X-CTRSNet simultaneously enables CT reconstruction and segmentation directly from the biplanar 2D X-ray images. For a fair comparison, we solely utilize the CT reconstruction branch (SPaDRNet module) while freezing the extra segmentation branch. We use the same Adam solver with X2CT-GAN to train X-CTRSNet for 200 epochs with a learning rate of 2e-4.

\begin{figure*}[t]
\centering
\includegraphics[width=\linewidth]{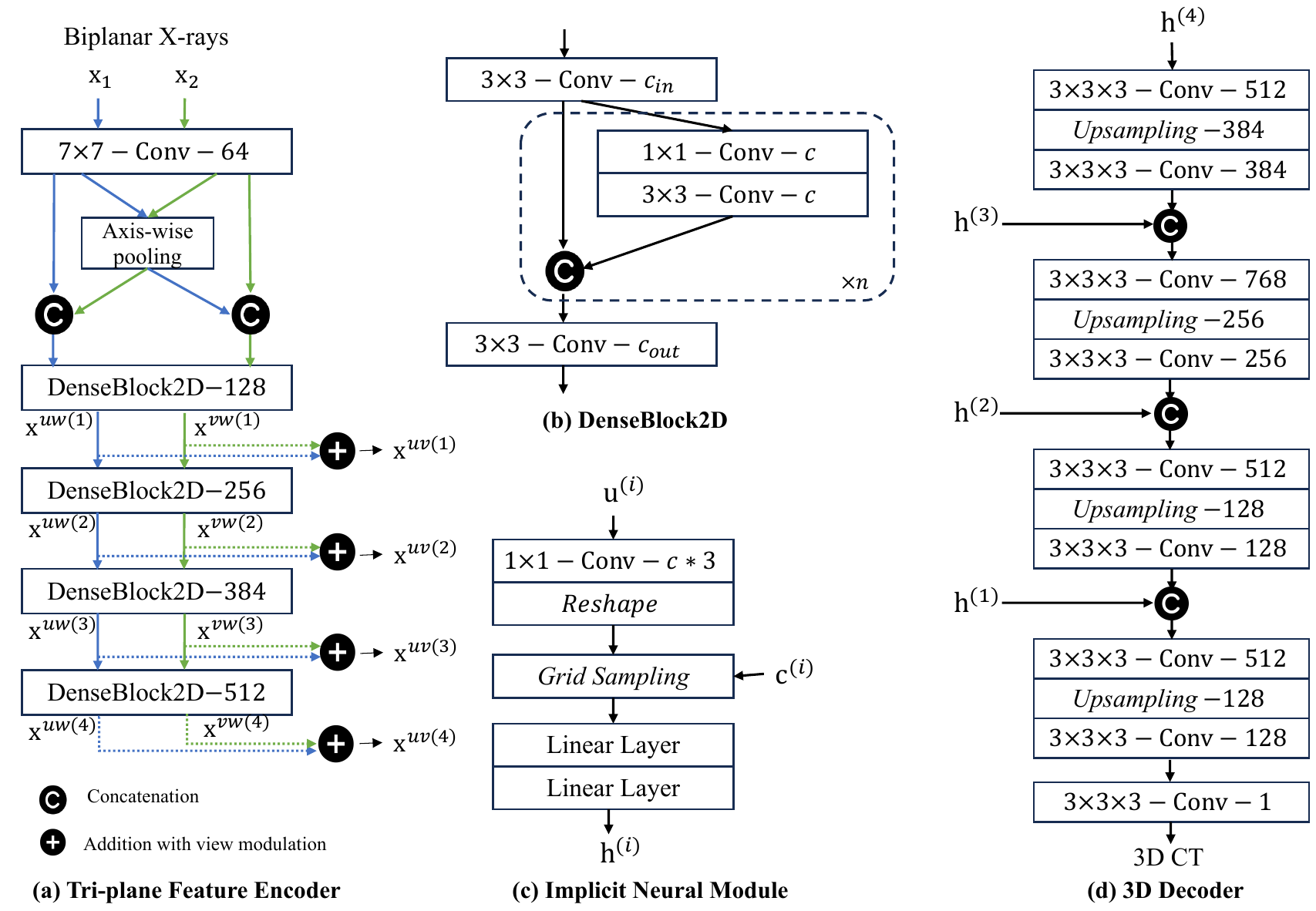} 
% \vspace{-4mm}
\caption{Detailed Model Architecture of each module in ICM-\textit{Reg}.}
\label{fig:tpnet}
\end{figure*}

\section{Model architecture of ICM-\textit{Reg}}
\label{tpnet}

Based on the implicit conditioning mechanism, we develop a new regression model, ICM-\textit{Reg}, by introducing a simple 3D decoder. 
As shown in Fig. \ref{fig:tpnet}, ICM-\textit{Reg} is comprised of a tri-plane decoupling generator, an implicit neural decoder, and a 3D decoder. The tri-plane decoupling generator takes the 2D biplanar X-rays $\mathbf{x}_1,\mathbf{x}_2$ as inputs. It then produces the multi-resolution tri-plane features ${\mathbf{x}^{uw,(i)}, \mathbf{x}^{vw,(i)}, i \in\{1, \cdots, N\}}$, and the fused feature planes ${\mathbf{x}^{uv,(i)}}$ modulated by the view embeddings $s_1=(1,0,1)$ and $s_2=(0,1,1)$. Note that we add a learnable embedding to ${\mathbf{x}^{uv,(4)}}$.
Right after that, three 2D lightweight convolutional decoders with shared weights input these feature planes to generate the tri-plane features ${\mathbf{u}^{(i)}, i \in\{1, \cdots, N\}}$ in a decoupling manner.
Then, ${\mathbf{u}^{(i)}}$ along with the assigned 3D coordinates $\mathbf{c}^{\left(i\right)}$ are fed into the implicit neural module to query the multi-resolution 3D structural features ${\mathbf{h}^{(i)}, i \in\{1, \cdots, N\}}$. Last, the 3D Decoder integrates ${\mathbf{h}^{(i)}}$ and upsamples them to obtain the final 3D CT. ICM-\textit{Reg} is directly optimized by a $\mathcal{L}_1$ loss function.

\section{Additional Results}
\label{result}

\subsection{Additional Qualitative Results}
\label{sec:qualitative}

\begin{figure*}[htbp]
  \centering
  \begin{subfigure}{0.95\linewidth}
    		\includegraphics[width=\linewidth]{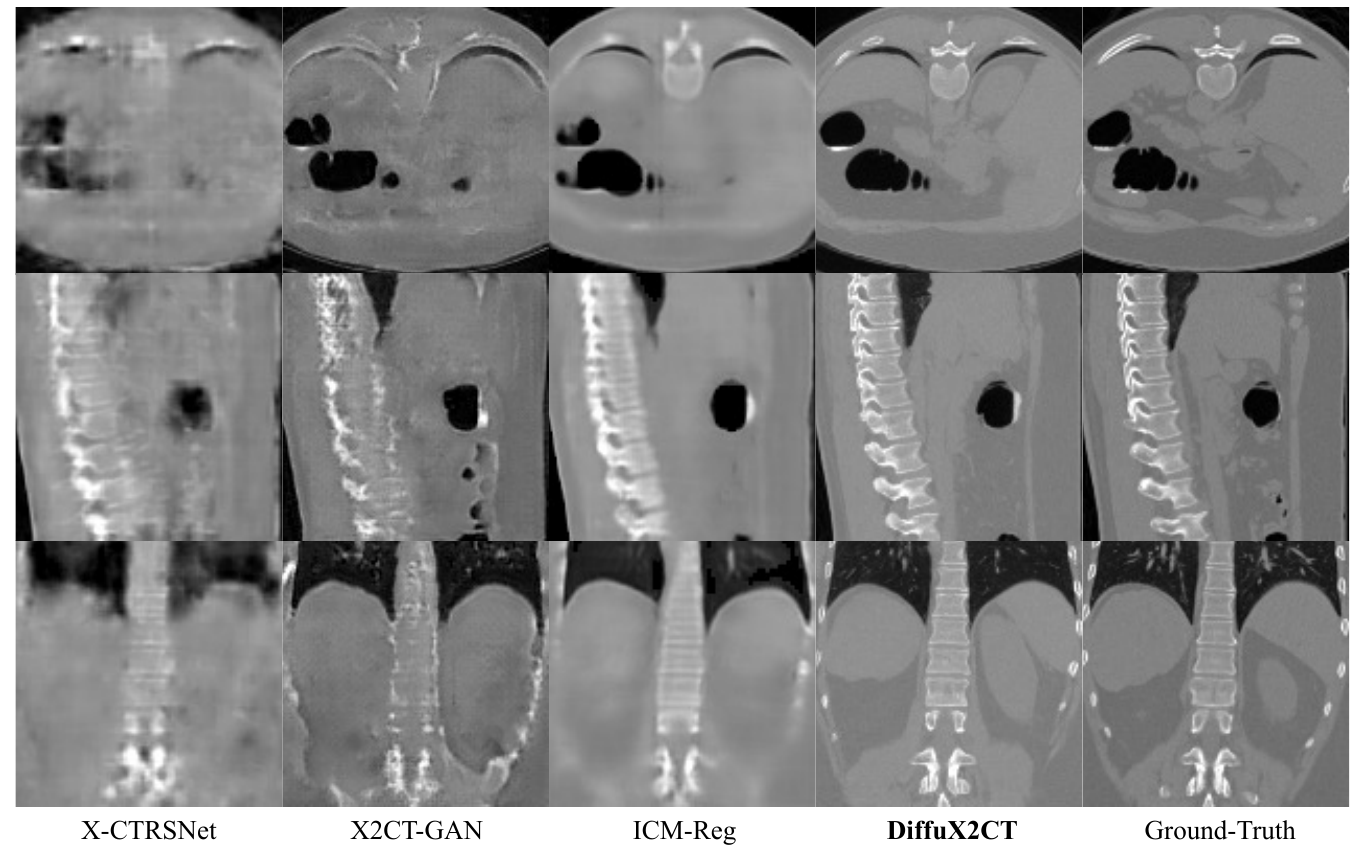}
    \caption{Case 1}
    \label{fig:short-a}
  \end{subfigure}
  \hfill
  \begin{subfigure}{0.96\linewidth}
    		\includegraphics[width=\linewidth]{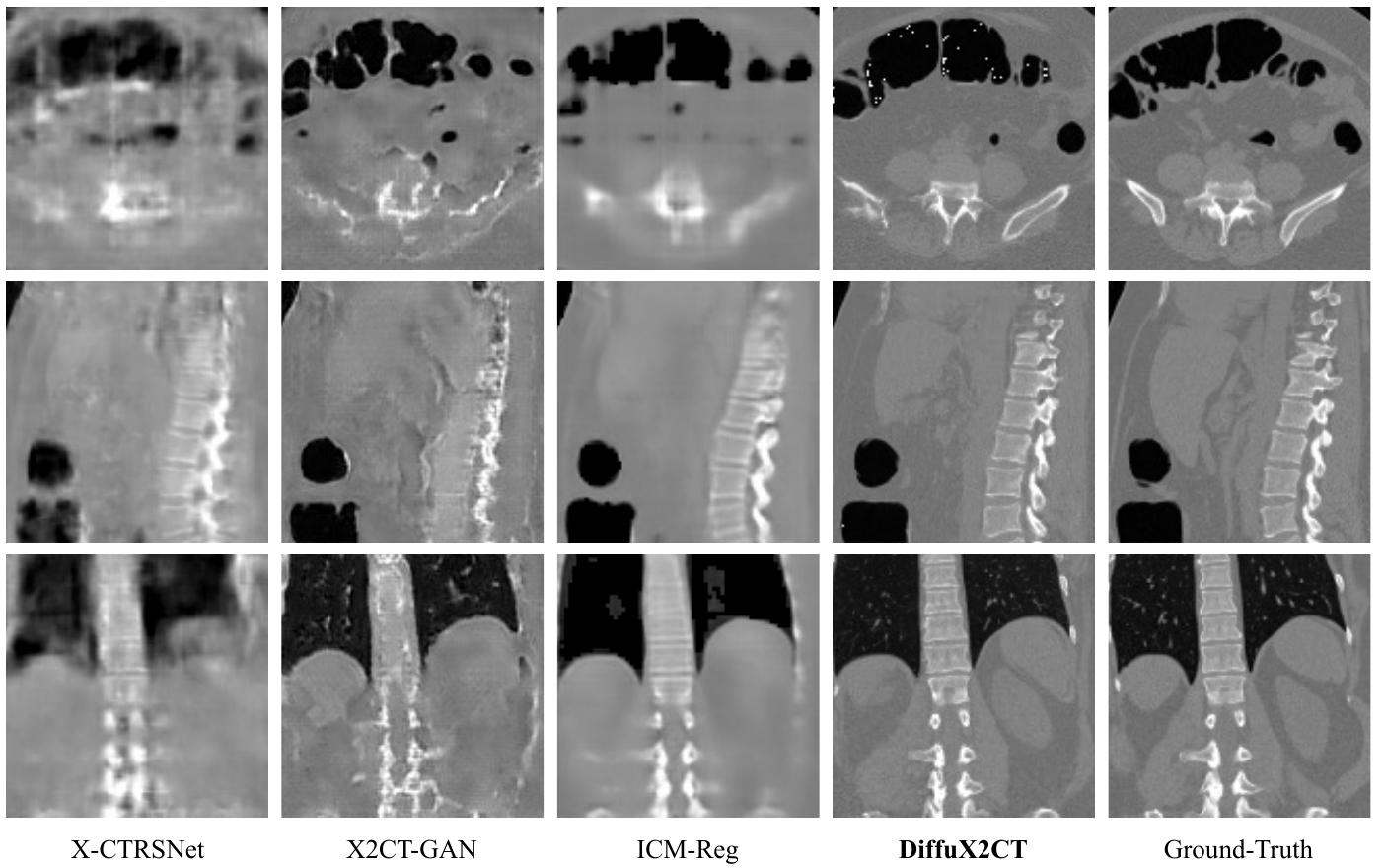}
    \caption{Case 2}
    \label{fig:short-b}
  \end{subfigure}
  \caption{Qulitative comparisons on the CTSpine1K datasets. The black part is the contrast agent. From top to bottom, we show the axial, sagittal, and coronal views. Best viewed by zoom in.}
  \label{fig:spine1}
\end{figure*}

\begin{figure*}[htbp]
  \centering
  \begin{subfigure}{0.96\linewidth}
    		\includegraphics[width=\linewidth]{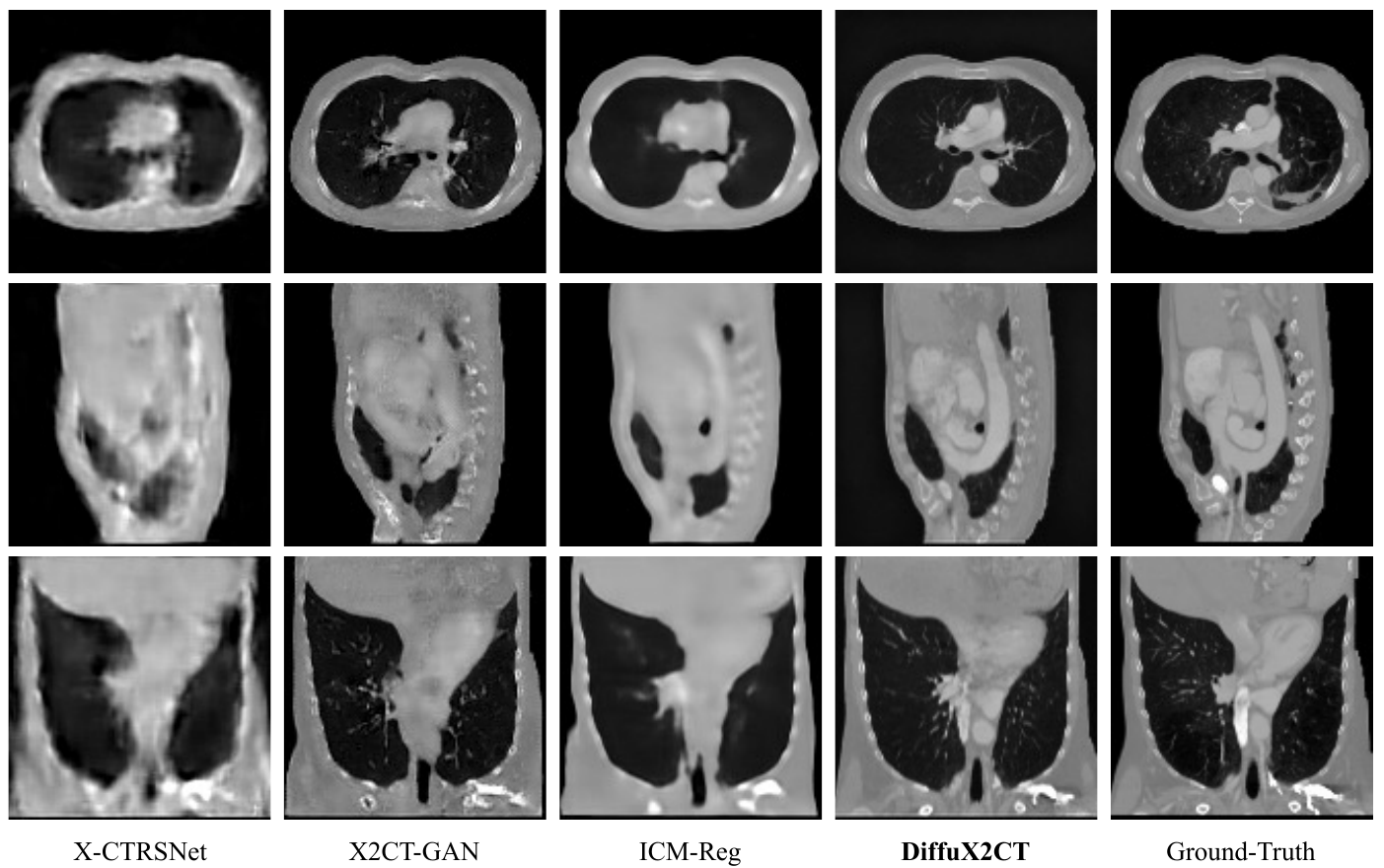}
    \caption{Case 1}
    \label{fig:short-a}
  \end{subfigure}
  \hfill
  \begin{subfigure}{0.95\linewidth}
    		\includegraphics[width=\linewidth]{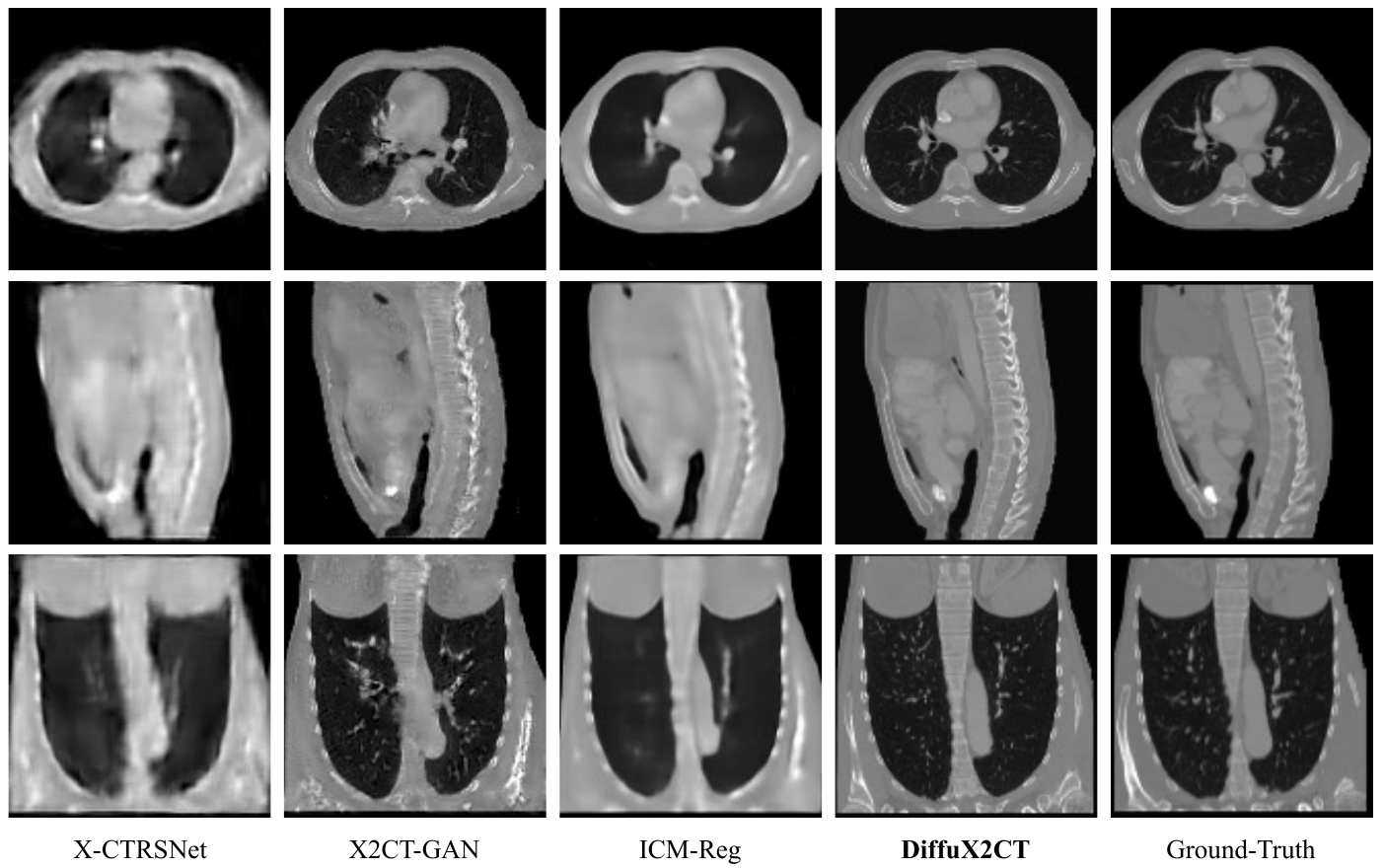}
    \caption{Case 2}
    \label{fig:short-b}
  \end{subfigure}
  \caption{Qulitative comparisons on the LIDC-IDRI datasets. The black part is the contrast agent. From top to bottom, we show the axial, sagittal, and coronal views. Best viewed by zoom in.}
  \label{fig:chest1}
\end{figure*}

\begin{figure*}[htbp]
  \centering
  \begin{subfigure}{0.95\linewidth}
    		\includegraphics[width=\linewidth]{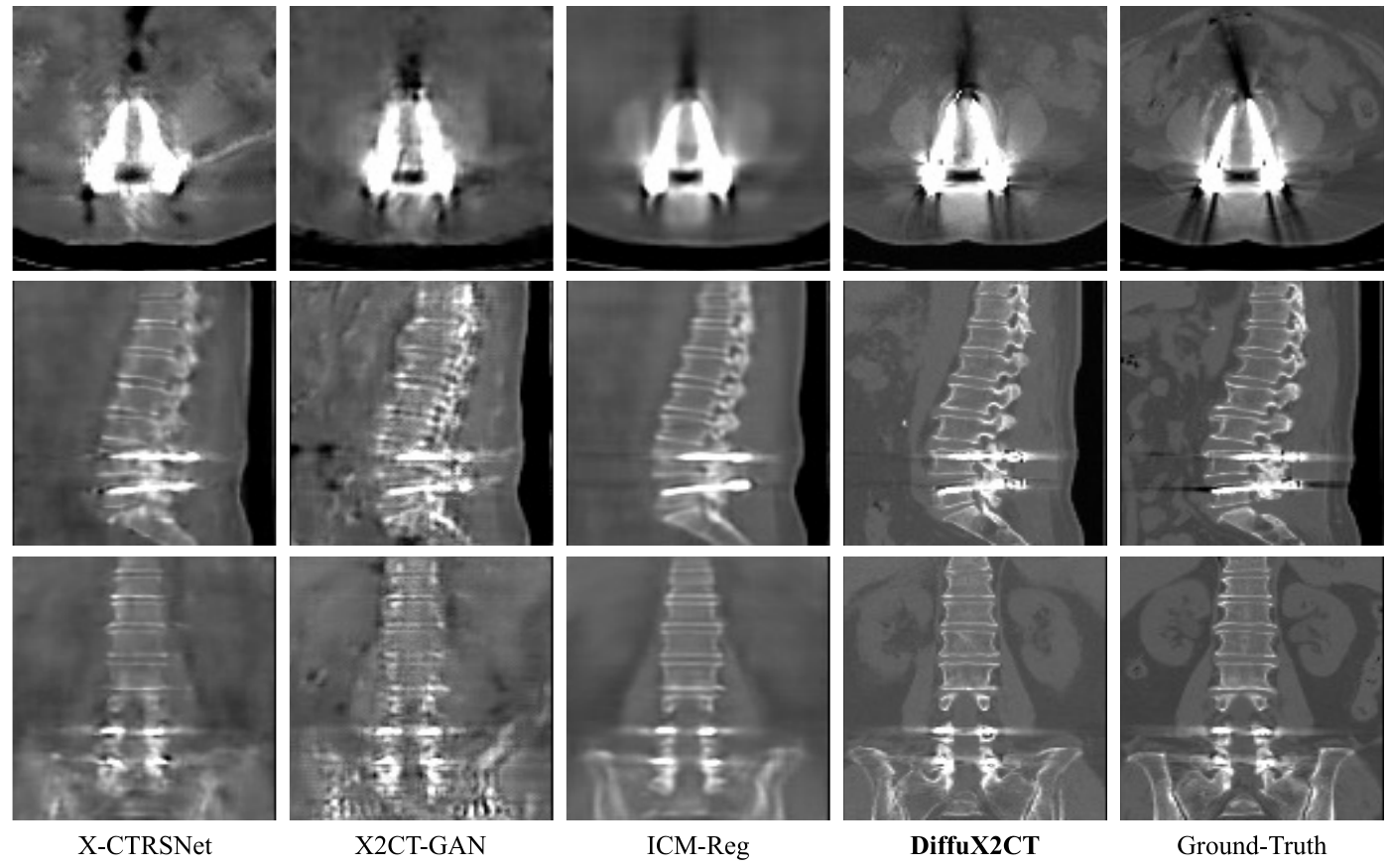}
    \caption{Case 1}
    \label{fig:short-a}
  \end{subfigure}
  \hfill
  \begin{subfigure}{0.95\linewidth}
    		\includegraphics[width=\linewidth]{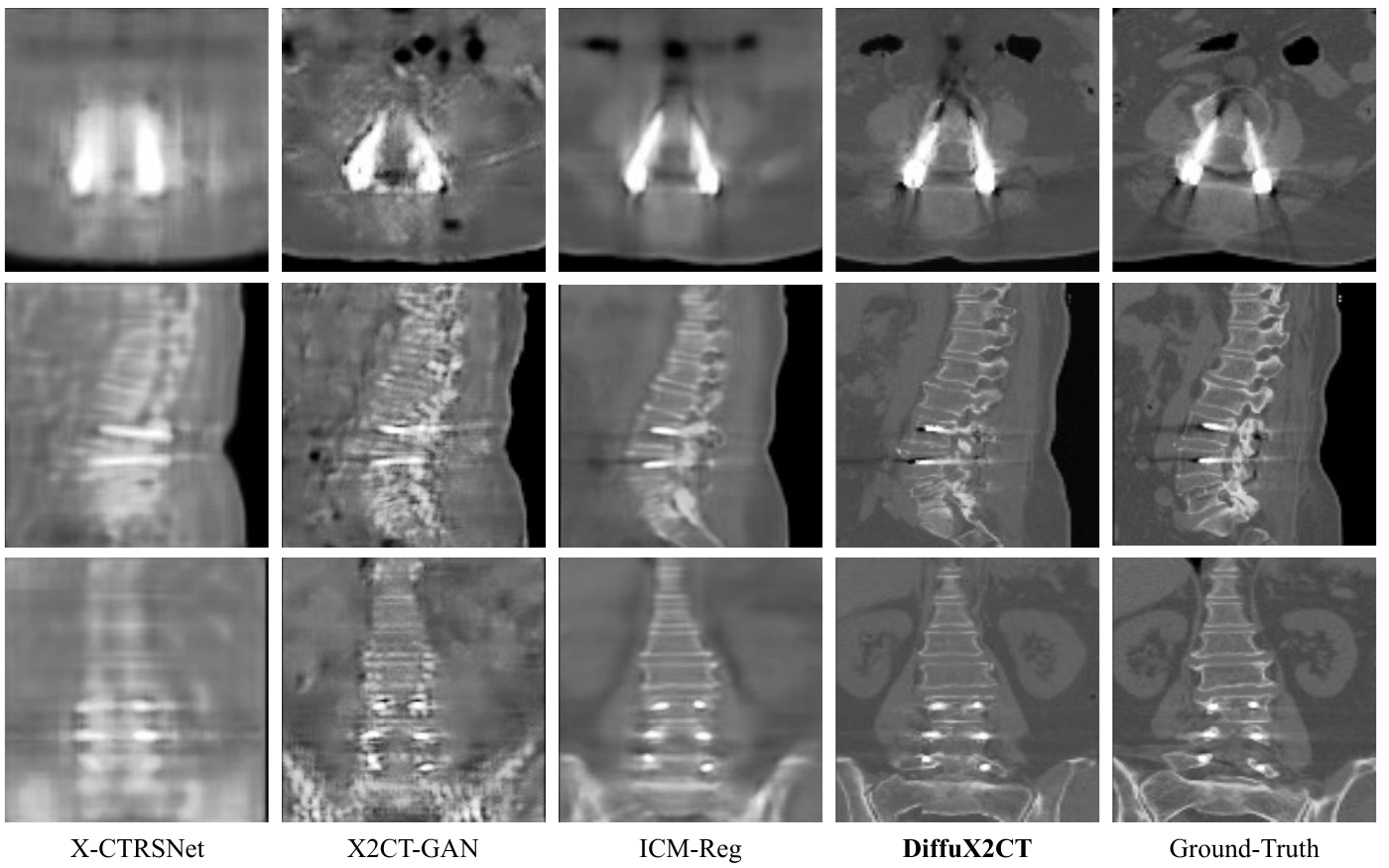}
    \caption{Case 2}
    \label{fig:short-b}
  \end{subfigure}
  \caption{Qulitative comparisons on the LumbarV datasets. The black part is the contrast agent. From top to bottom, we show the axial, sagittal, and coronal views. Best viewed by zoom in.}
  \label{fig:lumbar1}
\end{figure*}

\begin{figure*}[htbp]
  \centering
  \begin{subfigure}{0.95\linewidth}
    		\includegraphics[width=\linewidth]{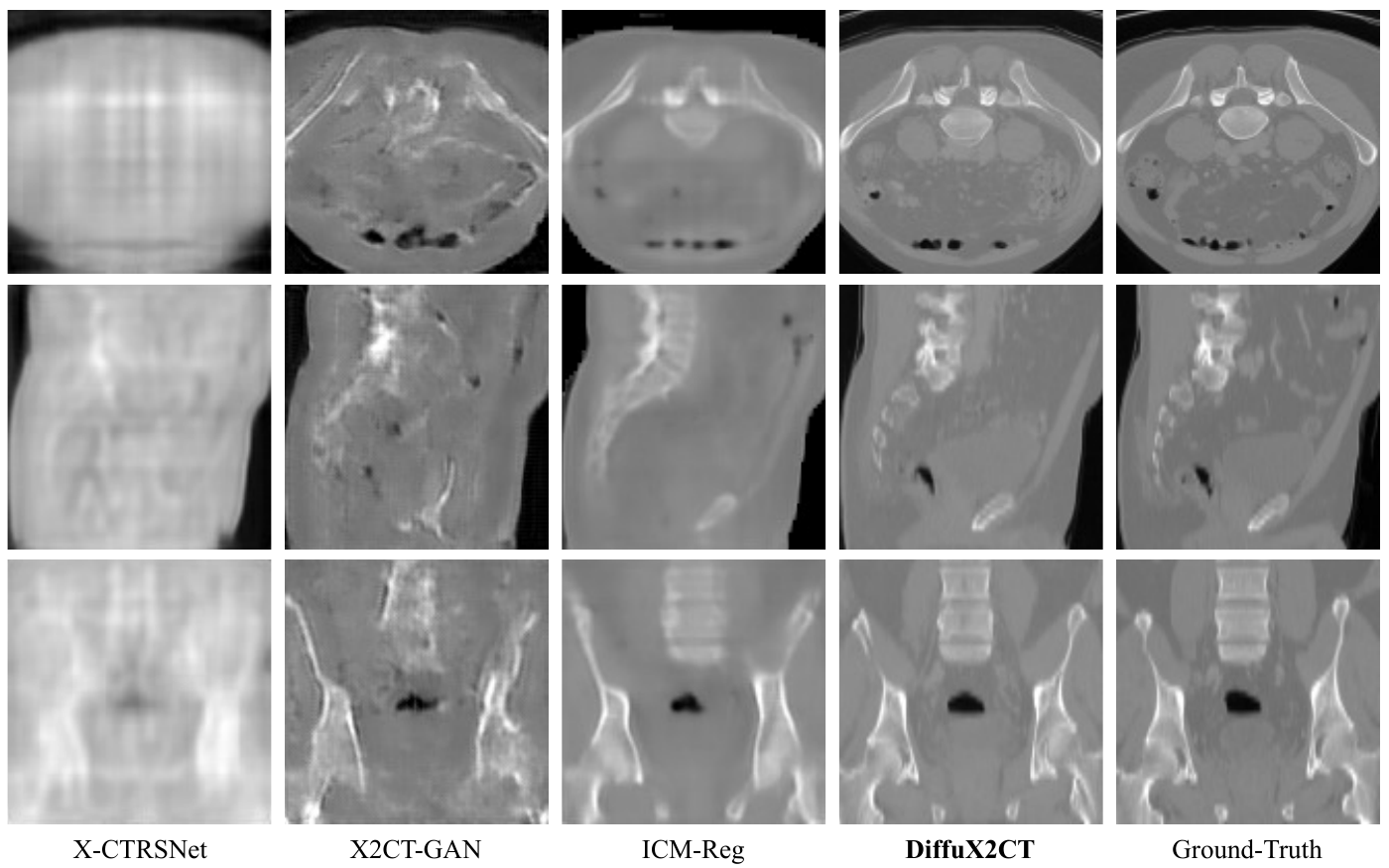}
    \caption{Case 1}
    \label{fig:short-a}
  \end{subfigure}
  \hfill
  \begin{subfigure}{0.95\linewidth}
    		\includegraphics[width=\linewidth]{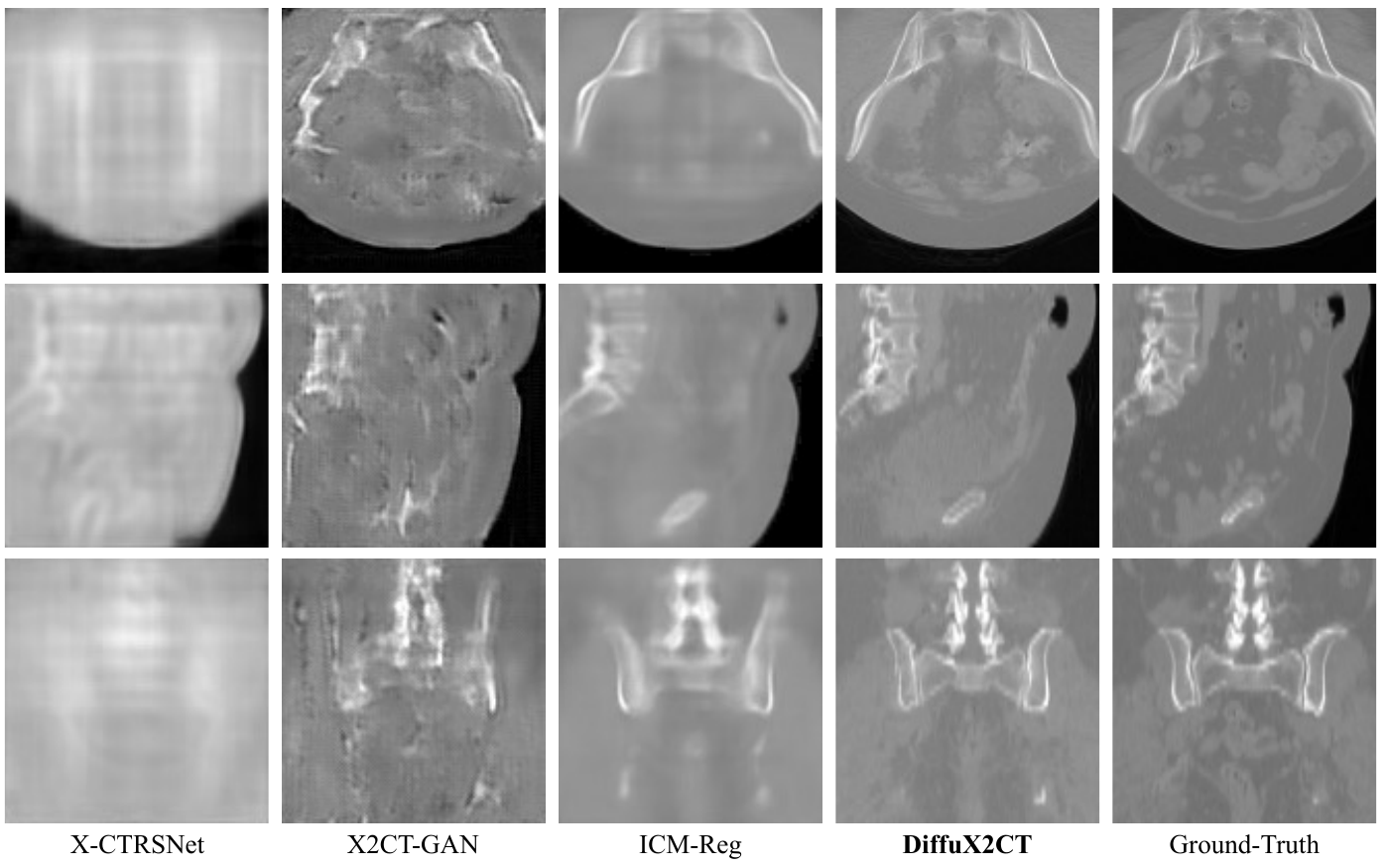}
    \caption{Case 2}
    \label{fig:short-b}
  \end{subfigure}
  \caption{Qulitative comparisons on the CTPelvic1K datasets. The black part is the contrast agent. From top to bottom, we show the axial, sagittal, and coronal views. Best viewed by zoom in.}
  \label{fig:pelvis1}
\end{figure*}

\begin{figure*}[htbp]
  \centering
  \begin{subfigure}{0.98\linewidth}
    		\includegraphics[width=\linewidth]{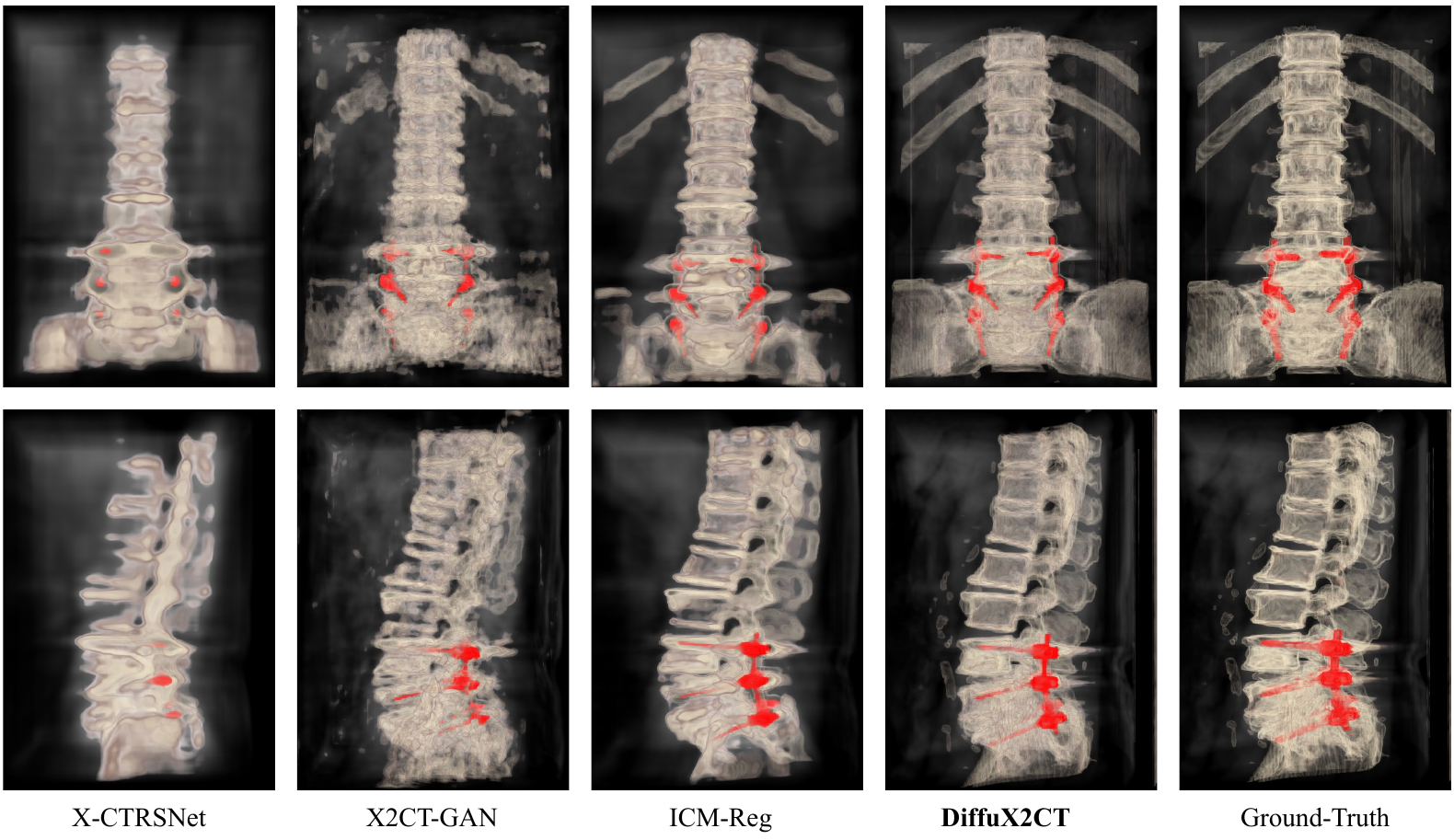}
    \caption{Case 1}
    \label{fig:short-a}
  \end{subfigure}
  \hfill
  \begin{subfigure}{0.98\linewidth}
    		\includegraphics[width=\linewidth]{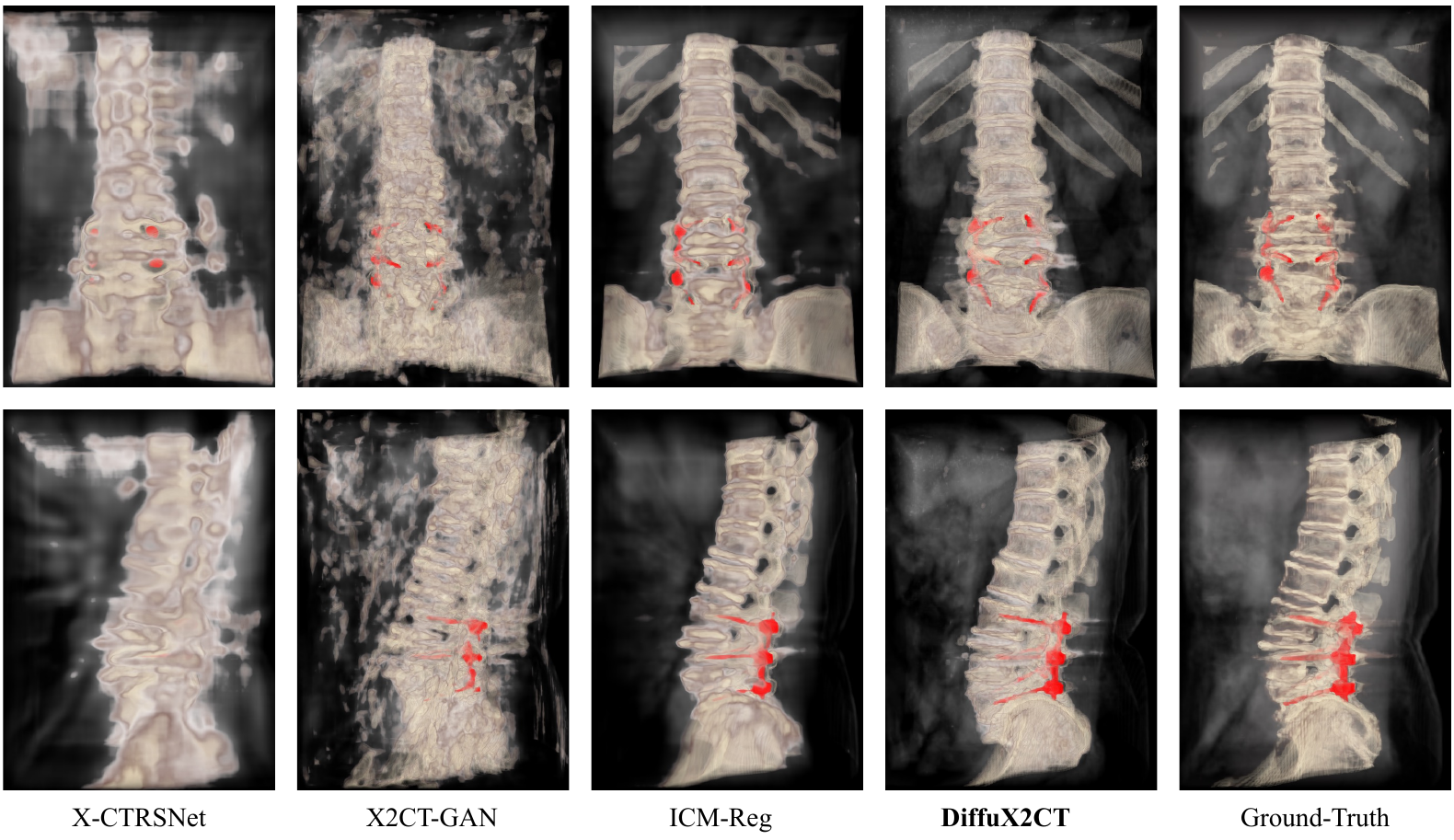}
    \caption{Case 2}
    \label{fig:short-b}
  \end{subfigure}
  \caption{Rendering results on the LumbarV datasets. We show the front view and the side view. Best viewed by zoom in.}
  \label{fig:lumbar2}
\end{figure*}

\begin{figure*}[htbp]
  \centering
  \includegraphics[width=\linewidth]{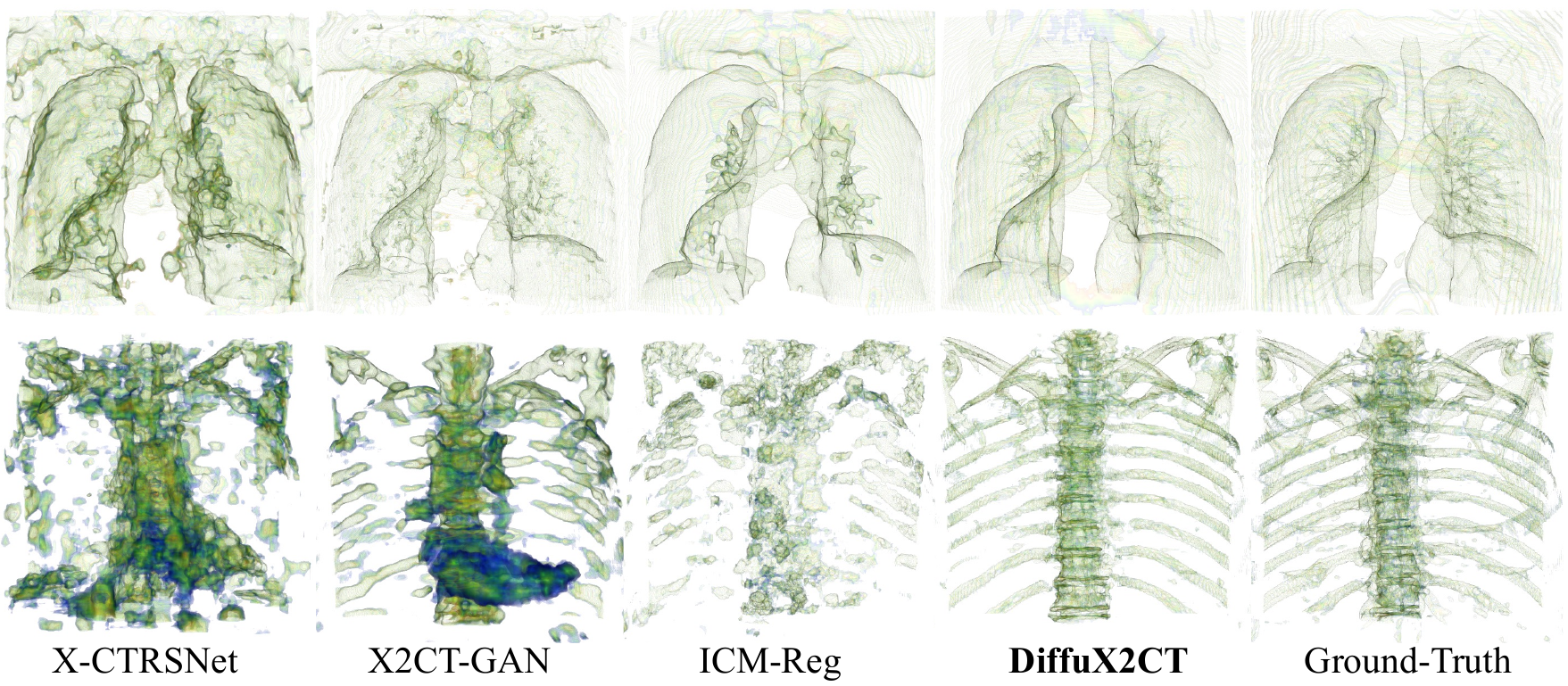}
  \caption{Rendering results on the CTSpine1K datasets. We show the visualizations of both soft tissues and bones. Best viewed by zoom in.}
  \label{fig:spine2}
\end{figure*}

\begin{figure*}[htbp]
  \centering
  \begin{subfigure}{0.99\linewidth}
    		\includegraphics[width=\linewidth]{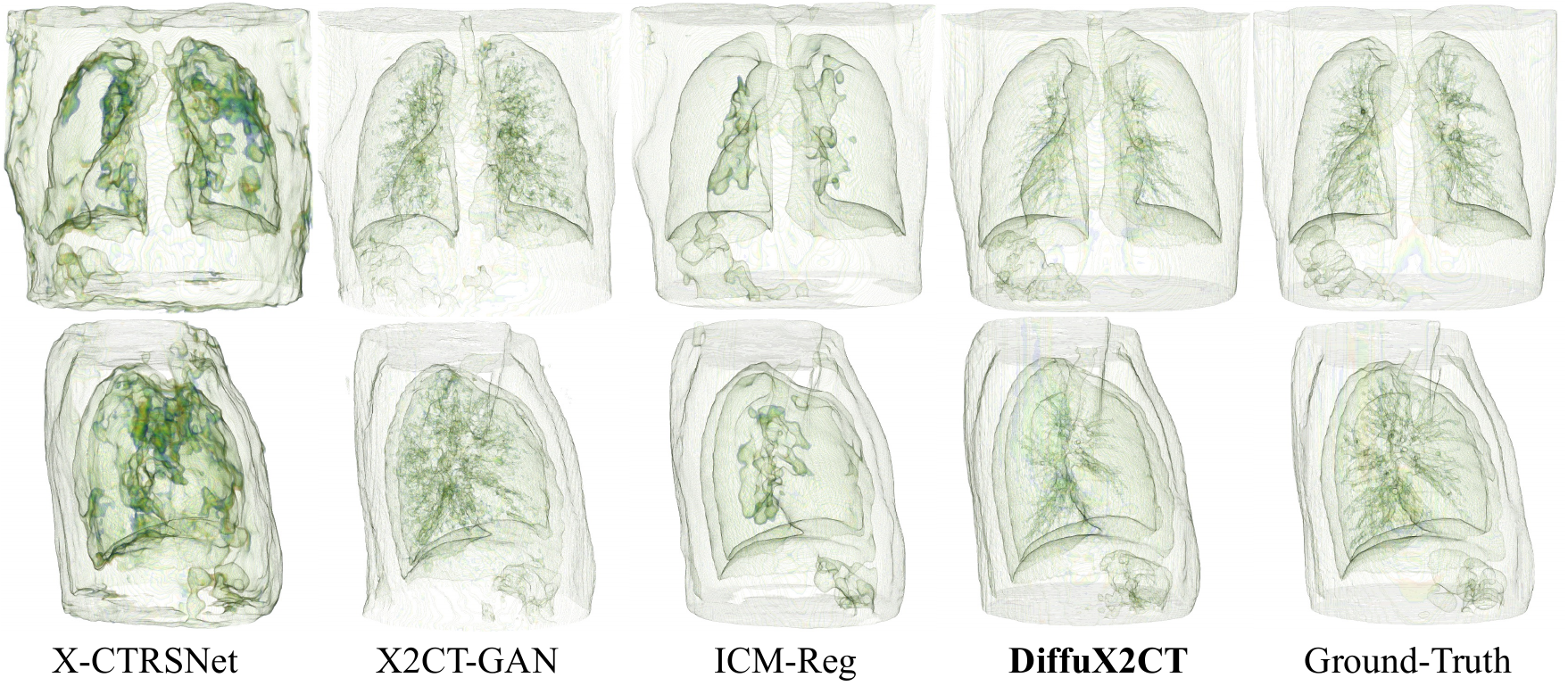}
    \caption{Case 1}
    \label{fig:short-a}
  \end{subfigure}
  % \hfill
  \begin{subfigure}{0.99\linewidth}
    		\includegraphics[width=\linewidth]{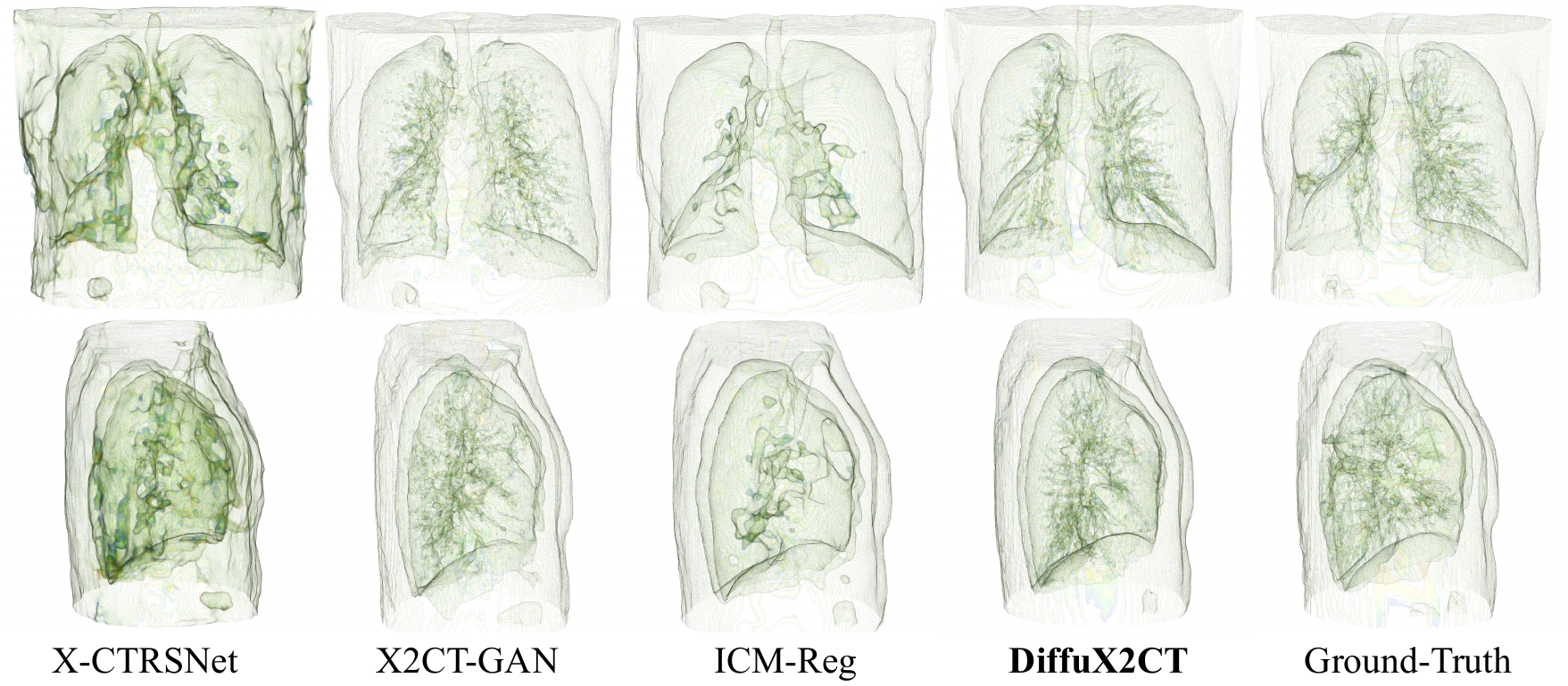}
    \caption{Case 2}
    \label{fig:short-b}
  \end{subfigure}
  \caption{Rendering results on the LIDC-IDRI datasets. We show the front view and the side view. Best viewed by zoom in.}
  \label{fig:chest2}
\end{figure*}

We first show the axial, sagittal, and coronal views of the reconstructed CT images on the four datasets in Fig. \ref{fig:spine1}, Fig. \ref{fig:chest1}, Fig. \ref{fig:lumbar1}, and Fig. \ref{fig:pelvis1}. 
Second, we provide the rendering results on the LumbarV dataset and highlight the implants with red color in Fig. \ref{fig:lumbar2}. It shows the structure of reconstructed bones and implants from the front and side perspectives. 
Moreover, we present the rendering results on the CTSpine1K and LIDC-IDRI datasets in Fig. \ref{fig:spine2} and Fig. \ref{fig:chest2}. For CTSpine1K, we adjust the shift value to visualize the soft tissues and bones, respectively. For LIDC-IDRI,
we show the front and side views like LumbarV.
It is evident that \textit{DiffuX2CT} surpasses other methods by a large margin in reconstructing high-quality CT images while recovering consistent structures with the ground-truth. In addition to generating bones and implants with accurate shapes, \textit{DiffuX2CT} can also reconstruct soft tissues well.

\subsection{Additional 3D Segmentation Results}
\label{seg}
\begin{figure*}[htbp]
  \centering
  \begin{subfigure}{0.999\linewidth}
    		\includegraphics[width=\linewidth]{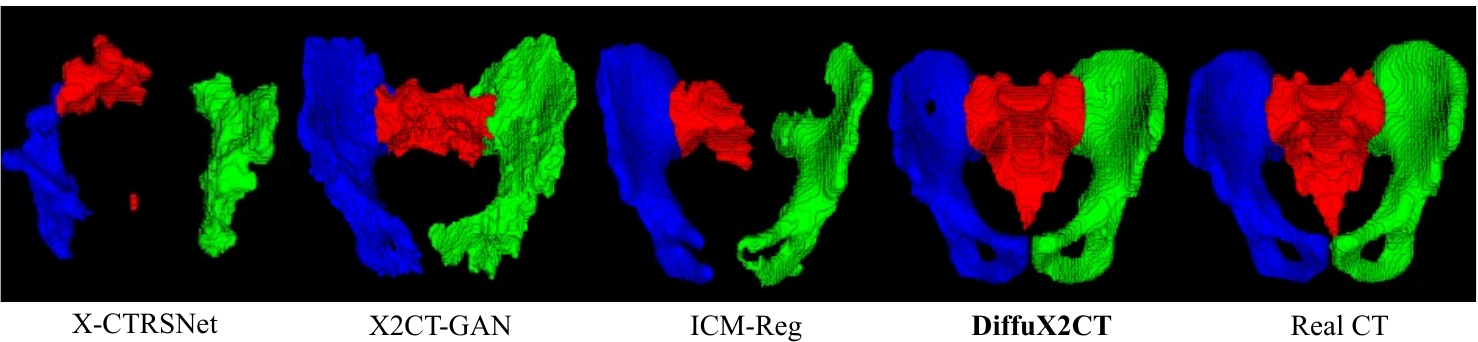}
    \caption{Case 1}
    \label{fig:short-a}
  \end{subfigure}
  \hfill
  \begin{subfigure}{0.999\linewidth}
    		\includegraphics[width=\linewidth]{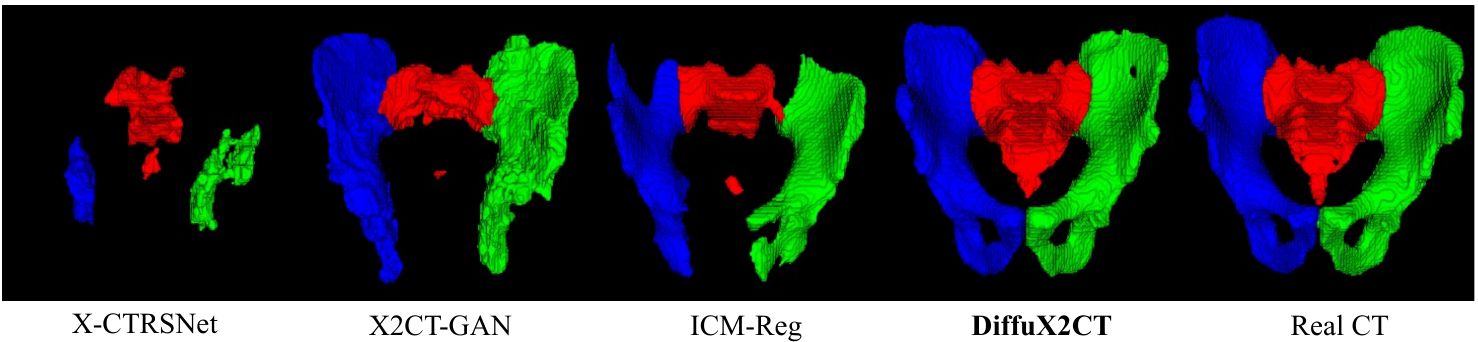}
    \caption{Case 2}
    \label{fig:short-b}
  \end{subfigure}
  \caption{The comparison of different reconstruction methods by rendering the 3D segmentation masks of the reconstructed CT images. Best viewed by zoom in.}
  \label{fig:seg}
\end{figure*}

To further demonstrate the effectiveness and practical significance of our \textit{DiffuX2CT}, we show comparison results of the reconstructed CT images in terms of 3D segmentation in Fig. \ref{fig:seg}, where ``Real CT'' represents the segmentation results of the real CT.
The segmentation masks of \textit{DiffuX2CT} display the highest resemblance to real CT images, whereas other methods yield inferior results with inaccurately reconstructed bone shapes. The 3D segmentation results further verify that our \textit{DiffuX2CT} is capable of reconstructing pelvic bones with accurate structures.
% \section{Additional Case Study}
% \label{case}

\bibliographystyle{splncs04}
\bibliography{main}